\documentclass[10pt,letterpaper,twocolumn]{article}

\usepackage[textwidth=504bp,textheight=9.5in,columnsep=24bp]{geometry}
\usepackage[T1]{fontenc}
\usepackage[version=3]{mhchem}
\usepackage{amsmath,amssymb}
\usepackage{newtxtext,newtxmath}
\usepackage{graphicx,xcolor}
\usepackage{microtype}
\usepackage{algorithm,algpseudocode}
\usepackage{tikz}
\usetikzlibrary{quantikz2}
\usepackage{makecell,booktabs,multirow,array}
\usepackage[font=small,labelfont=bf]{caption}
\usepackage{authblk}
\usepackage[numbers,sort&compress,super]{natbib}

\usepackage[hidelinks]{hyperref}

\title{\textbf{Combining Hardware Efficient Ansatz and Spin-based Quantum Phase Estimation For Reliable Molecular Excited-State Calculations}}
\author[1]{Young Kyun Ahn}
\author[1]{Young Min Rhee}
\affil[1]{Department of Chemistry, Korea Advanced Institute of Science and
  Technology, Daejeon 34141, Korea}
\affil[ ]{Correspondence: \texttt{ymrhee@kaist.ac.kr}}
\date{}

\expandafter\let\expandafter\npjwidefigure
  \csname figure*\endcsname
\expandafter\let\expandafter\endnpjwidefigure
  \csname endfigure*\endcsname

\NewCommandCopy{\npjincludegraphics}{\includegraphics}
\NewDocumentCommand{\npjcolumnimage}{O{}m}
  {\npjincludegraphics[width=\columnwidth]{#2}}

\ExplSyntaxOn
\RenewDocumentEnvironment{figure}{O{tbp}+b}
  {
    \clist_map_inline:nn
      {fig:hardware_spin_profiles_theory,fig:hardware_four_count_fraction,fig:linear_fitting,fig:orth_check_numbers}
      {
        \str_if_in:nnT {#2} {\label{##1}}
          {\cs_set_eq:NN \includegraphics \npjcolumnimage \clist_map_break:}
      }
    \bool_set_false:N \l_tmpa_bool
    \clist_map_inline:nn {fig:graphs_asym,fig:graphs_sym_s2}
      {
        \str_if_in:nnT {#2} {\label{##1}}
          {\bool_set_true:N \l_tmpa_bool \clist_map_break:}
      }
    \bool_if:NTF \l_tmpa_bool
      {\begin{npjwidefigure}[tp]#2\end{npjwidefigure}}
      {\begin{npjcolumnfigure}[#1]#2\end{npjcolumnfigure}}
  }
  {}
\RenewDocumentEnvironment{equation}{+b}
  {
    \str_if_in:nnT {#1} {\begin{split}} {\small}
    \begin{npjequation}#1\end{npjequation}
  }
  {}
\ExplSyntaxOff

\begin{document}
\twocolumn[{%
  \begin{@twocolumnfalse}
    \maketitle
    \begin{abstract}


We propose a spin-sensing strategy that combines a $z$-component spin ($\hat S_z$) preserving ansatz and quantum phase estimation of its transverse spin $\hat S_{x}$. The discrete spectrum of $\hat S_{x}$ enables probabilistic discrimination of the total-spin values without direct measurements with a quadratic operator $\hat S^{2}$ to reduce the overall circuit burden.
By applying the strategy to variational quantum deflation calculations of \ce{LiH} and \ce{BeH2} through statevector simulations, we demonstrate improvements in the separation of singlet and triplet excited states as well as reductions in the number of lower-lying states that enter the deflation penalty.
We also devise a calibration scheme to extract the total spin populations from an arbitrary state with a fixed $S_z$ value. We test the calibration scheme on an IBM Nighthawk quantum processor to demonstrate the spin-sector populations can indeed be reconstructed with reasonable reliability even from noisy occupation-counting measurements. Although the reconstruction accuracy decreases with the circuit depth, the modest ancilla requirements and the shallowness of the constructed circuits offer a promising route toward spin-selective quantum algorithms on near-term devices.

    \end{abstract}
    \vspace{1em}
  \end{@twocolumnfalse}
}]

\section{Introduction}

The variational quantum eigensolver (VQE) \cite{peruzzoVariationalEigenvalueSolver2014, mccleanTheoryVariationalHybrid2016, tillyVariationalQuantumEigensolver2022, cerezoVariationalQuantumAlgorithms2021, caoQuantumChemistryAge2019, mcardleQuantumComputationalChemistry2020} is a hybrid quantum-classical algorithm that finds optimal solutions of certain problems by variationally treating a parameterized quantum circuit. It is particularly suitable for noisy intermediate-scale quantum (NISQ) devices \cite{preskillQuantumComputingNISQ2018, bhartiNoisyIntermediatescaleQuantum2022, kokailSelfverifyingVariationalQuantum2019, hempelQuantumChemistryCalculations2018, googleaiquantumandcollaboratorsHartreeFockSuperconductingQubit2020, chenDemonstrationAdiabaticVariational2020, harriganQuantumApproximateOptimization2021, paganoQuantumApproximateOptimization2020, kandalaHardwareefficientVariationalQuantum2017}, and has been proposed as an amenable solution for a range of problems such as electronic structure calculations \cite{omalleyScalableQuantumSimulation2016, kandalaHardwareefficientVariationalQuantum2017}. Because direct energy minimization targets the ground state, additional strategies are needed to access excited states. Variational quantum deflation (VQD)\cite{higgottVariationalQuantumComputation2019} finds successive states by adding overlap penalties against previously converged states, whereas subspace-search VQE (SSVQE)\cite{nakanishiSubspacesearchVariationalQuantum2019} optimizes several mutually orthogonal states using a shared unitary transformation. In VQD, the number of overlap constraints grows with the number of previously computed states, and errors in lower-lying solutions can affect subsequent optimizations. Identifying the intended excited state becomes particularly challenging when different spin sectors remain accessible within the same variational space\cite{tillyVariationalQuantumEigensolver2022, higgottVariationalQuantumComputation2019}.

Whichever tactic is adopted for modelling excited states with a VQE-based approach, an assumption of the many-electron wavefunction, or an ansatz, needs to be made. While there are ways to form the ansatz such that only physically acceptable wavefunctions are considered during the optimization processes, such ans\"atze are often avoided due to their unattractive circuit properties. For example, unitary coupled-cluster (UCC) ans\"atze \cite{shenQuantumImplementationUnitary2017, romeroStrategiesQuantumComputing2018} are guaranteed to generate wavefunctions that are correct in terms of particle numbers and many other properties but they require very deep circuits with heavy resource requirements. Alternatively, qubit-space methods such as the qubit coupled cluster (QCC)\cite{ryabinkinQubitCoupledCluster2018} and its iterative variant (IQCC)\cite{ryabinkinIterativeQubitCoupled2020} reduce circuit depth by operating directly in the qubit space, but ensuring physical validity of the resulting states still requires careful design. Thus, many forms of hardware efficient ansatz (HEA) have been proposed to reduce the circuit requirements \cite{kandalaHardwareefficientVariationalQuantum2017}. However, optimized HEA may correspond to a state that violates physical validity especially when applied to excited states, and thus care must be taken to avoid such an artifact. Thus, in electronic structure theory with quantum computing, obtaining physically meaningful excited states with VQE requires more than simply reproducing their energies alone. Namely, one needs to make sure that the computed states possess the correct quantum-mechanical characters. Two most prominent characters of any molecular systems will be the total number of electrons in the system, and the spin state of each electronic state that one considers. In variational excited-state calculations with HEA, enforcing such properties is often nontrivial.

At least, keeping the number of electrons correct can be achieved with relative ease through the use of symmetry-preseving ansatz \cite{gardEfficientSymmetrypreservingState2020, bistafaAccuracyPotentialHardwareEfficient2025}. Correctly preserving the spin state is more demanding as keeping the numbers of electrons separately in spin $\alpha$ and spin $\beta$ subspaces, namely preserving $\hat{n}_{\alpha}$ and $\hat{n}_{\beta}$, is not enough.
Representative approaches to spin-control within variational frameworks include ansatz-level symmetry adaptation \cite{gardEfficientSymmetrypreservingState2020}, explicit observable-based penalties \cite{lyuSymmetryEnhancedVariational2023}, projector-based symmetry restoration \cite{sekiSymmetryadaptedVariationalQuantum2020}, and constraint-based excited-state formulation \cite{gochoExcitedStateCalculations2023a}. For example, Gard et al. proposed symmetry-preserving ans\"atze in which hyperspherical parameters are fixed \textit{a priori} to confine the variational search to a subspace of a definite total spin \cite{gardEfficientSymmetrypreservingState2020}. On the other hand, Lyu et al. incorporated the expectation value of $\hat{S}^{2}$ as a penalty term to enforce total spin selectivity \cite{lyuSymmetryEnhancedVariational2023}. However, these approaches become excessively complex as the system size grows or at least incur an $\mathcal{O}(n^{2})$-scale measurement overhead from the Pauli string decomposition of $\hat{S}^{2}$. 

In this work, we take a different route. Rather than encoding all constraints into an ansatz of a deep circuit or enforcing measurements of an unfavorable number of additional observables, we combine a symmetry-preserving ansatz with a lightweight, ancilla-assisted discriminator \cite{abramsQuantumAlgorithmProviding1999, xieOrthogonalStateReduction2022}. Specifically, building on the symmetry-preserving ansatz, which was originally formulated as a particle-number conserving and hardware-efficient circuit under Jordan-Wigner encoding \cite{jordanWignerPauliExclusion1928}, we construct an $\hat{S}_{z}$ conserving variant by fixing $\hat{n}_{\alpha}$ and $\hat{n}_{\beta}$. This $\hat{S}_{z}$ conserving symmetry preserving (SSP) ansatz yields a variational search space that is chemically meaningful and yet NISQ-compatible.
To avoid introducing substantial measurement overhead incurred by large Pauli decompositions associated with explicit evaluation on $\hat{S}^{2}$, and to further control the spin multiplicity, we introduce a shallow phase-estimation-like routine with respect to spin axis rotation by $\hat{S}_{x}$ that encodes spin-sector information into a small ancilla register. When we ignore spin-orbit coupling, namely when we assume a spin-free electronic Hamiltonian, $\hat{S}_{x}$ commutes with $\hat{H}$, and the ancilla readout can act as a probabilistic spin-sector discriminator without altering the Hamiltonian estimation procedures. In shot-based settings, samples flagged as out of the pursued spin sector can be penalized or discarded early, effectively suppressing spin contamination \cite{lowdinQuantumTheoryManyParticle1955,stahlQuantifyingReducingSpin2022} during optimization. We demonstrate the utility of this strategy in numerical VQD calculations of molecular excited states. Combining the SSP ansatz with $\hat{S}_{x}$-based screening improves the separation of different spin multiplets during sequential excited-state optimization.

The practical use of this screening procedure also depends on whether its spin information remains accessible under hardware noise. We therefore examine the spin readout on an IBM Nighthawk processor. Although noise makes individual outcomes unreliable for direct spin screening, the full readout distributions retain spin-dependent features under matched circuit and hardware conditions. Using reference states with mathematically well-defined spin characters, we propose a calibration scheme with which the spin-state distribution of a circuit with an arbitrary parameter set can be reconstructed from noisy measurement results. Through the actual hardware application, we show reconstruction remains effective up to moderate circuit depths even with present day's noisy machine.

\section{Methods}
\subsection{Variational quantum deflation}

For completeness, let us first briefly overview VQD, which is a sequential excite-state extension \cite{higgottVariationalQuantumComputation2019} of VQE. Basically, if we assume that we know the ground electronic state given by $|\psi^{(0)}\rangle$ and $k-1$ additional lowest electronic states $|\psi^{(j)}\rangle$ ($j = 0, \dots, k-1$), we can find the $k$-th excited state  $|\psi^{(k)} \rangle$ by minimizing the augmented cost function
\begin{equation}
\mathcal{L}_{k}(\Theta) = \langle\psi(\Theta)|\hat{H}|\psi(\Theta)\rangle
+ \sum_{j=0}^{k-1}{c_{j}\langle\psi(\Theta)|\psi^{(j)}\rangle\langle\psi^{(j)}|\psi(\Theta)\rangle}
\end{equation}
where $c_{j}$ is a penalty weight that enforces orthogonality of the solution to $|\psi^{(j)} \rangle$. The weights must be chosen on the order of the excitation energies. If they are too small, the optimizer collapses onto a previously found state, while if they are too large, the penalty terms amplify floating-point noise on the cost function. In fact,
among various excited-state VQE methods, here we adopt VQD for our benchmark as its sequential structure makes it particularly susceptible to the accumulation of numerical errors. Namely, each new state requires orthogonality checks against all previously converged states, and the number of penalty terms grows linearly with the excitation index $k$, which is known to induce error accumulations \cite{tillyVariationalQuantumEigensolver2022}. By restricting the variational search to a specific spin subspace and thereby reducing the number of states that must be orthogonalized against, the procedures may become more resilient to noise. Yet, we note that the spin-selective screening introduced later in this report is not limited to VQD and can also be straightforwardly applied to other excited-state VQE schemes such as SSVQE.

\subsection{Symmetry preserving ansatz}

The symmetry preserving (SP) ansatz \cite{gardEfficientSymmetrypreservingState2020, bistafaAccuracyPotentialHardwareEfficient2025, anselmettiLocalExpressiveQuantumnumberpreserving2021, barkoutsosQuantumAlgorithmsElectronic2018} is a hardware-efficient, particle-number conserving circuit widely used in NISQ implementations. It consists of a repetitive two-qubit rotation block $A(\theta, \phi)$, which couples $\alpha$ and $\beta$ spin states associated with the same spatial orbitals. The circuit $A(\theta, \phi)$ is given by
\begin{equation}
A\left(\theta,\phi\right) =
\begin{bmatrix}
1 & 0 & 0 & 0 \\
0 & \cos\theta & e^{i\phi}\sin\theta & 0 \\
0 & e^{-i\phi}\sin\theta & -\cos\theta & 0 \\
0 & 0 & 0 & 1
\end{bmatrix},    \label{eq:SP}
\end{equation}
which can be implemented with three CNOT gates (Figure \ref{fig:Agate}). Repeating this block as shown in Figure \ref{fig:SPansatz} through all qubit lanes yields the full SP circuit.

\begin{figure}
    \centering
    \begin{quantikz}[align equals at=1.5]
        & \gate[2]{A\left(\theta,\phi\right)} & \\
        &&
    \end{quantikz}
    $\equiv$
    \begin{quantikz}[align equals at=1.5]
        & \targ{}   &                                               & \ctrl{1}  &                                   & \targ{}   & \\
        & \ctrl{-1} & \gate{R\left(\theta,\phi\right)^{\dagger}}    & \targ{}   & \gate{R\left(\theta,\phi\right)}  & \ctrl{-1} & 
    \end{quantikz}
    \caption{Circuit representation of $A\left(\theta, \phi\right)$, where $R\left(\theta, \phi\right) = R_{z}\left(\phi+\pi\right)R_{y}\left(\theta+\pi/2\right)$.}
    \label{fig:Agate}
\end{figure}

\begin{figure}
    \centering
    \begin{quantikz}
        \lstick{$\ket{1}_{1\alpha}$} & \gate[2]{A\left(\theta_{1},\phi_{1}\right)}\gategroup[5, steps=2, style={dashed,rounded corners}, label style={label position=above}]{$\mathbf{SP}(\Theta,\Phi)$}\gategroup[5,steps=3, style={draw=none}, label style={label position=below right, yshift=-0.2cm}]{$\times M$}  & & \\
        \lstick{$\ket{1}_{1\beta}$}  &                                                & \gate[2]{A\left(\theta_{2},\phi_{2}\right)}       & \\
        \lstick{$\ket{1}_{2\alpha}$} & \gate[2]{A\left(\theta_{3},\phi_{3}\right)}    &                                                   & \\
        \lstick{$\ket{\cdots}$}      &                                                & \gate[2]{A\left(\theta_{4},\phi_{4}\right)}       & \\
        \lstick{$\ket{0}_{n\beta}$}  &                                                &                                                   &
    \end{quantikz}
    \caption{Composite quantum circuit of the SP ansatz that conserves the total number of particles.}
    \label{fig:SPansatz}
\end{figure}

This SP ansatz conserves the number of $|1\rangle$, and thus under Jordan-Wigner encoding \cite{jordanWignerPauliExclusion1928}, it conserves the total number of particles over a domain the ansatz chooses to mix. For example, applying the $A$-gate separately to the $\alpha$- and the $\beta$-qubits ensures the conservation of $\langle \hat{S}_{z} \rangle$.
Thus, to enforce $\hat{S}_{z}$ symmetry more strongly, we construct an $\hat{S}_{z}$-conserving variant SSP by applying SP operations independently on the registers that correspond to $\alpha$ and $\beta$ spin-orbitals, thereby fixing $n_\alpha$ and $n_\beta$. The scheme is shown in Figure~\ref{fig:SSPansatz}, and we can understand that this ansatz trivially conserves $\langle S_{z} \rangle$, because $m_{z}$ is given as $(n_\alpha-n_\beta)/2$. In addition, owing to the reduced number of connections, the dimensionality of the problem reduces from $C(2n, 2k)$ to $C(n,k+m_{z}) \cdot C(n, k-m_{z})$ with $n$ spatial orbitals and $2k$ active electrons, where $C(\cdot,\cdot)$ denotes the conventional binomial combinations.
In Figure~\ref{fig:SSPansatz}, we have additionally included a diagonal two-qubit phase gate to supplement the missing relative-phase degrees of freedom between relevant basis components that are suppressed when $\alpha$ and $\beta$ registers are treated independently. More specifically, because separately applying SP to ${\alpha}$ and ${\beta}$ subspaces cannot generate relative phases between the singly-occupied basis components $\ket{1_\alpha0_\beta}$ and $\ket{0_{\alpha}1_{\beta}}$, and because such relative phases are essential for correctly representing states across different spin multiplets, the $P\left(\xi_{1}, \xi_{2}, \xi_{3}\right)$ gate defined as
\begin{equation}
    P(\xi_{1},\xi_{2},\xi_{3}) =
    \begin{bmatrix}
    1 & 0 & 0 & 0 \\
    0 & e^{i\xi_{1}} & 0 & 0 \\
    0 & 0 & e^{i\xi_{2}} & 0 \\
    0 & 0 & 0 & e^{i(\xi_{1}+\xi_{2}+\xi_{3})}
    \end{bmatrix} 
\end{equation}
is included to restore these missing degrees of freedom without breaking the conservation condition of $\langle\hat{S}_{z}\rangle$. Note that the overall ansatz given in Figure~\ref{fig:SSPansatz} only requires a ladder-like topology of qubit entanglements, where each $\alpha$-$\beta$ pair forms a local two-qubit interaction, with the pairs connected as a chain as shown in Figure~S1 in Supplementary Information (SI).

\begin{figure}
\centering
\resizebox{\columnwidth}{!}{
    \begin{quantikz}
    \lstick{$\ket{1}_{1\alpha}$}& \gate[3,]{A(\theta_{1},\phi_{1})} \gateinput{$\bullet$}\gateoutput{$\bullet$}\gategroup[7,steps=2, style={inner sep=1pt}]{$\mathbf{SP}_{\alpha}$}\gategroup[8,steps=6, style={rounded corners, inner xsep=10pt, inner ysep=13pt, dashed, xshift=4pt, yshift=7pt}, label style={label position=below right, xshift=25pt,yshift=-0.2cm}]{$\times M$}\gategroup[8,steps=6,style={draw=none, inner xsep=10pt, inner ysep=13pt, dashed, xshift=4pt, yshift=7pt},label style={label position=above}]{$\mathbf{SSP}(\Theta, \Phi, \Xi)$} & & & & \ctrl[wire style={"\xi_{1}"}]{1}\gategroup[8, steps=2, style={inner xsep=4pt, xshift=2pt}]{Phase} & \phase{\scriptstyle\xi_{5}} & & \\
    \lstick{$\ket{1}_{1\beta}$}& & & \gate[3,]{A(\theta_{4},\phi_{4})} \gateinput{$\bullet$} \gateoutput{$\bullet$} \gategroup[7,steps=2, style={inner sep=1pt}]{$\mathbf{SP}_{\beta}$} & & \control{} & \phase{\scriptstyle\xi_{6}} & & \\
    \lstick{$\ket{1}_{2\alpha}$}& \gateinput{$\bullet$}\gateoutput{$\bullet$} & \gate[3,]{A(\theta_{3},\phi_{3})} \gateinput{$\bullet$}\gateoutput{$\bullet$} & & & \ctrl[wire style={"\xi_{2}"}]{1} & \phase{\scriptstyle\xi_{7}} & & \\
    \lstick{$\ket{1}_{2\beta}$}& & & \gateinput{$\bullet$}\gateoutput{$\bullet$} & \gate[3,]{A(\theta_{6},\phi_{6})} \gateinput{$\bullet$}\gateoutput{$\bullet$} & \control{} & \phase{\scriptstyle\xi_{8}} & & \\
    \lstick{$\ket{\cdots}$}& \gate[3,]{A(\theta_{2},\phi_{2})} \gateinput{$\bullet$}\gateoutput{$\bullet$} & \gateinput{$\bullet$}\gateoutput{$\bullet$} & & & \ctrl[wire style={"\xi_{3}"}]{1} & \phase{\scriptstyle\xi_{9}} & & \\
    \lstick{$\ket{\cdots}$}& & & \gate[3,]{A(\theta_{5},\phi_{5})} \gateinput{$\bullet$}\gateoutput{$\bullet$} & \gateinput{$\bullet$}\gateoutput{$\bullet$} & \control{} & \phase{\scriptstyle\xi_{10}} & & \\
    \lstick{$\ket{0}_{n\alpha}$}& \gateinput{$\bullet$}\gateoutput{$\bullet$} & & & & \ctrl[wire style={"\xi_{4}"}]{1} & \phase{\scriptstyle\xi_{11}} & & \\
    \lstick{$\ket{0}_{n\beta}$}& & & \gateinput{$\bullet$}\gateoutput{$\bullet$} & & \control{} & \phase{\scriptstyle\xi_{12}} & &
    \end{quantikz}
    }
    \caption{Construction of the SSP ansatz with the SP ansatz gates applied separately on $\alpha$ and $\beta$ spin domains followed by phase gates for providing up to three different phases to three different occupations. Bullets denote the input and the output qubits of the corresponding gates.}
    \label{fig:SSPansatz}
\end{figure}
\subsection{Discriminating spin states through measurements with ancilla qubits}

Quantum phase estimation (QPE) is a quantum algorithm that estimates the eigenphases of a unitary operator using ancilla qubits. In general, eigenphases are real-valued and cannot be represented exactly with a finite number of binary digits. Estimating them to higher precision therefore requires a larger ancilla register or additional iterations in an iterative scheme. In contrast, the eigenvalues of a spin component in a spin system are integer- or half-integer-valued, and when the spin eigenvalues are treated with phase estimation, they may be mapped to distinct eigenphases represented by a finite and relatively small number of ancilla qubits. This naturally requires a rotation operation, and it can also be constructed with a circuit based on Figure \ref{fig:Agate}. As QPE requires controlled powers of unitary operators, efficiently implementing the operations will be important for keeping the circuit depth manageable, and the finiteness of the required spin analyzing operation will surely help in regard to this depth issue.

With this argument in mind, we propose a two-step scheme for screening spin states. First, the SSP ansatz fixes $m_{z}$ to the target spin quantum number $S_{\mathrm{tgt}}$, restricting the variational state to total-spin sectors with $S\ge S_{\mathrm{tgt}}$. Second, we measure $\hat{S}_{x}$ and identify undesired outcomes satisfying $|m_{x}|>S_{\mathrm{tgt}}$, which arises only from higher-spin sectors with $S> S_{\mathrm{tgt}}$. Some screening weight is then incorporated into the VQD scheme to suppress such undesired off-target spin contributions during excited-state optimization.

Throughout this work, we consider a nonrelativistic spin-free electronic Hamiltonian in the absence of spin--orbit coupling and external magnetic fields. The Hamiltonian is therefore invariant under global spin rotations and satisfies
\begin{equation}
    [\hat{H},\hat{S}_{a}]=0,
    \qquad a\in\{x,y,z\},
\end{equation}
as well as $[\hat{H},\hat{S}^{2}]=0$. We define the transverse spin rotation
\begin{equation}
    U_{x}(\theta)=e^{i\theta\hat{S}_{x}},
    \label{eq:Ux}
\end{equation}
and because $\hat{H}$ and $\hat{S}_{x}$ commutes, any transverse rotation leaves the energy expectation value invariant:
\begin{equation}
    \bra{\psi}
    U_{x}^\dagger(\theta)\hat{H}U_{x}(\theta)
    \ket{\psi}
    =
    \langle\psi | \hat{H} | \psi \rangle.
\end{equation}

The SSP ansatz explained in the previous part restricts the variational state to a sector with a fixed $m_{z}$, but it does not uniquely fix the total spin quantum number $S$. In general, within the manifold dictated by the SSP ansatz, states with different $S$ may coexist with the same $m_{z}$, causing spin contamination. Rather than explicitly evaluating $\hat S^{2}$, which typically leads to large Pauli decompositions and substantial measurement overhead, we employ transverse-spin readout based on $U_{x}(\theta)$ in Eq.\ \ref{eq:Ux} to probabilistically obtain spin-sector information.
On the basis of $|S,m_{x}\rangle$, $U_{x}(\theta)$ is diagonal and acts purely to encode the phase as
\begin{equation}
U_{x}(\theta)\,|S,m_{x}\rangle = e^{i m_{x} \theta}\,|S,m_{x}\rangle.
\label{eq:phase_encoding}
\end{equation}
Thus, a shallow phase-estimation routine \cite{kitaevQuantumMeasurementsAbelian1995a, nielsenQuantumComputationQuantum2010, aspuru-guzikSimulatedQuantumComputation2005} on $U_{x}(\theta)$ 
can be devised to return the value of $m_{x}$ (Figure \ref{fig:R_{x}_screening}). Because a state with a fixed $m_{z}$ can be expanded as
\begin{equation}
    |S,m_{z}\rangle= \sum_{m_{x}} d^{S}_{m_{x},m_{z}} \, |S,m_{x} \rangle
    \label{eq:wigner_expansion}
\end{equation}
with the Wigner $d$-matrix coefficients $d^{S}_{m_{x},m_{z}}$ \cite{wignerEinigeFolgerungenAus1927a, sakuraiModernQuantumMechanics2011}, the probability that a measurement of $\hat{S}_x$ on $|S,m_{z}\rangle$ yields a result of $|m_{x}|\le |m_{z}|$ is given by
\begin{equation}
P(|m_x|\le|m_z|;\;|S,m_{z}\rangle)=\sum_{m_{x}=-|m_{z}|}^{|m_{z}|}\left|d^S_{m_{x},m_{z}}\right|^2 .
\label{eq:pass_probability}
\end{equation}
For the sector with smallest $S$, namely the sector with $S=|m_{z}|$, the pass probability is unity. For higher-spin sectors it is generally smaller than unity and exhibits an $S$-dependent, including parity-dependent, variation as shown in Table~\ref{tbl:screeningprobability}.

\begin{figure}
    \centering
    \resizebox{\columnwidth}{!}{
    \begin{quantikz}
        \lstick[4]{$\ket{\psi_{0}}$}& \gate[4]{\mathbf{SSP(\Theta,\Phi,\Xi)}} \slice[style = {transform canvas={xshift=-3pt}, black}]{$|\Psi_{1}\rangle$} & \gate[2]{\hat{R}_{x}\left(\frac{\pi}{2}\right)}\gategroup[6,steps=4,style={rounded corners, dashed},label style={label position=above, yshift=3pt}]{$\mathcal{QPE}_{\hat{S}_{x}}$} & & \gate[2]{\hat{R}_{x}\left(\frac{\pi}{4}\right)} & \slice[style={transform canvas={xshift=3pt}, black}]{$|\Psi_{2}\rangle$} & \slice[style=black]{$|\Psi_{3}\rangle$}&\\
        & & & & & & &\\
        & & & \gate[2]{\hat{R}_{x}\left(\frac{\pi}{2}\right)} & & \gate[2]{\hat{R}_{x}\left(\frac{\pi}{4}\right)} & &\\
        & & & & & & &\\
        \lstick[2]{$\ket{0}_{m_{x}}$}& \gate{H} & \ctrl{-3} & \ctrl{-1} & & & \gate[2]{\mathcal{QFT}^{\dagger}}&\meter[2]{m_{x}}\\
        & \gate{H} & & & \ctrl{-4} & \ctrl{-2} & &
    \end{quantikz}}
    \caption{Scheme of screening statevector with controlled-$\hat{R}_{x}$ operators. }
    \label{fig:R_{x}_screening}
\end{figure}

\begin{table}
    \caption{Exemplary values of $P\left(|m_{x}|\leq |m_{z}|;\; |S, m_{z}\rangle \right)$).}
    \label{tbl:screeningprobability}
    \centering
    \begin{tabular}{l*{4}{>{\centering\arraybackslash}p{1.6cm}}}
    \toprule
    \multirow{2}{*}{$S$} & \multicolumn{4}{c}{$m_{z}$} \\
    \cmidrule(l){2-5}
     & $0$ & $1$ & $2$ & $3$ \\
    \midrule
    $0$ & $1$      &           &            &          \\
    $1$ & $0$      & $1$       &            &          \\
    $2$ & $1/4$    & $1/2$     & $1$        &          \\
    $3$ & $0$      & $7/32$    & $13/16$    & $1$      \\
    $4$ & $9/64$   & $9/32$    & $11/32$    & $15/16$  \\
    $5$ & $0$      & $1/8$     & $539/2048$ & $305/512$\\
    \bottomrule
    \end{tabular}
    \label{table:P_S_mz}
\end{table}

Let us now focus on how our phase estimation scheme works. For this, let us first trace the state evolution through the circuit in Figure~\ref{fig:R_{x}_screening}. The circuit is composed of a Jordan-Wigner domain in the upper part and an ancilla domain with $n_\mathrm{anc}$ qubits in the lower part. Assuming that the initial state of the molecule of our interest in the Jordan-Wigner domain is an eigenstate $\ket{\psi_1}=\ket{S, m_{z}}$, the total initial state $\ket{\Psi_1}$ shown in the figure including the ancilla domain can be written as
\begin{equation}
    \ket{\Psi_{1}} \equiv \ket{\psi_{1}}\otimes \ket{+}_{n_{\text{anc}}}
    = \ket{S, m_{z}} \otimes \ket{+}_{n_{\text{anc}}} ,
\end{equation}
where the ancillas are prepared in the Hadamard state
\begin{equation}
  \ket{+}_{n_{\text{anc}}} = \sum_{j=-2^{n_{\text{anc}}-1}}^{2^{n_{\text{anc}}-1}-1}{\frac{1}{\sqrt{2}^{n_{\text{anc}}}}\ket{j}}
\end{equation}
as again displayed in Figure~\ref{fig:R_{x}_screening}. Here, we have adopted signed integers for assigning the range for $j$. This is more convenient than the convention with unsigned integers as it is a more natural choice for encoding both non-negative and negative values of $m_{x}$ into the ancilla register. Thus, we can write the initial state as
\begin{equation}
    \ket{\Psi_{1}}
    ={\left( \sum_{m_{x}=-S}^{S}{d_{m_{x},m_{z}}^{S}\ket{S,m_{x}}} \right)}\otimes\left\{\sum_{j=-2^{n_{\text{anc}}-1}}^{2^{n_{\text{anc}}-1}-1}{\frac{1}{\sqrt{2}^{n_{\text{anc}}}}\ket{j}}\right\} .
\end{equation}
Then, the state after the phase estimation circuit, $\ket{\Psi_2}$, becomes
\begin{equation}
    \begin{split}
        \ket{\Psi_{2}}
        &=\mathcal{QPE}_{\hat{S}_{x}} \sum_{m_{x}=-S}^{S}{\frac{d_{m_{x},m_{z}}^{S}}{\sqrt{2}^{n_{\text{anc}}}}\cdot \left\{\sum_{j=-2^{n_{\text{anc}}-1}}^{2^{n_{\text{anc}}-1}-1}\left\{\ket{S,m_{x}}{\ket{j}}\right\}\right\}} \\
        &=\sum_{m_{x}=-S}^{S}{\frac{d_{m_{x},m_{z}}^{S}}{\sqrt{2}^{n_{\text{anc}}}}\cdot \left\{\sum_{j=-2^{n_{\text{anc}}-1}}^{2^{n_{\text{anc}}-1}-1}\left\{e^{2\pi i j m_{x}/2^{n_{\text{anc}}} } \ket{S,m_{x}}{\ket{j}}\right\}\right\}} .
    \end{split}
\end{equation}
In the above equation, we have used the fact that QPE is a linear operator and that $\ket{S,m_{x}}$ is an eigenstate of each rotation in QPE with the eigenvalue specified in the equation.
Note that in Figure~\ref{fig:R_{x}_screening} and in the above equation we define $\mathcal{QPE}_{\hat{S}_{x}}$ only with the controlled-rotation stages by excluding the initial Hadamard layer and the final inverse quantum Fourier transform (QFT). Finally, after the inverse QFT stage, the state becomes
\begin{equation}
    \ket{\Psi_{3}}
        =\sum_{m_{x}=-S}^{S}{\left\{d_{m_{x},m_{z}}^{S}\cdot\ket{S, m_{x}}\otimes\ket{m_{x}}\right\}} .
\end{equation}
Before the ancilla measurement, the system and ancilla registers remain entangled as shown in the above equation. Measuring the ancilla register yields one value of $m_{x}$ and conditionally projects the system register onto the corresponding $\hat S_{x}$ eigenspace.

While the screening based on single axis rotation with $\hat{S}_{x}$ already provides useful spin discrimination, its effectiveness may degrade for higher spin multiplicities. As can be seen from Table~\ref{tbl:screeningprobability}, the pass probability $P(S, m_{z})$ for contaminating states becomes non-negligible when the target spin multiplicity is high. In this case, the discrimination can be enhanced by applying identical phase-estimation scheme along a second spin axis (Figure~S2). Because $\hat{S}_{x}$ and $\hat{S}_{y}$ do not commute, the Jordan-Wigner domain will collapse to any one eigenvector of $\hat{S}_{y}$ with its corresponding eigenvalue encoded on the second ancilla domain. The total spin quantum number $S$ is preserved by these measurements, and if needed we can in principle repeat this scheme with alternating spin axes until the screening probability meets a desired level.

\begin{figure}
    \centering
    \includegraphics[width=\linewidth]{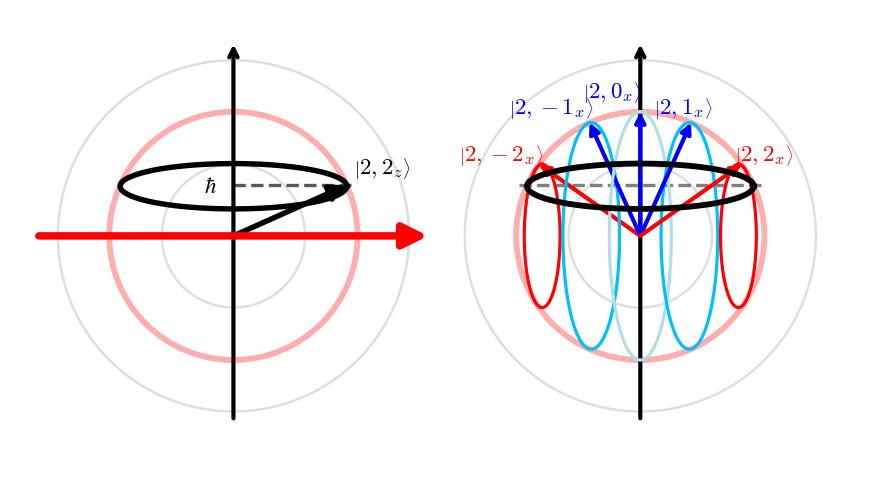}
    \caption{Schematic illustration of the $\hat{S}_{x}$-based phase estimation. When there is a spin state $\ket{S,m_{z}}$, shown in this case with $\ket{2,2}$ on the left, it is in general a superposition of $\hat S_{x}$-eigenstates with $\ket{S,m_{z}}=\sum_{m_{x}} c_{m_{x}}\ket{S,m_{x}}$. Rotation by $U(\theta)=e^{i\hat S_{x}\theta}$ tags each component with a phase $e^{i m_{x}\theta}$, and standard QPE registers the estimates of $m_{x}$ on the ancilla. When a state with $S$ is targeted, (1) an initial state is set to $m_{z} = S$, (2) the state is then propagated through the ansatz circuit, and (3) invalid states with $|m_{x}| > S$ are marked for filtering out or for penalization.}
    \label{fig:sxqpe}
\end{figure}

Our approach can be schematically explained with Figure~\ref{fig:sxqpe} illustrating layers of spin spheres with different $S$ values. Here, a state with a fixed $m_{z}$ value will correspond to a ring and its size will depend on the $S$ value as in the left panel. When looked at from the $x$-direction, this state is a superposition of multiple rings with multiple $m_{x}$ values as shown on the right. Upon time-evolution by a rotation with $\hat{S}_{x}$, these components with different $m_{x}$ values will precess with different frequencies, and the actual $m_{x}$ value can be deduced based on the frequency. In general, SSP may generate any state with $S \geq |m_{z}|$ and it again is a superposition of multiple states from the viewpoint of $m_{x}$. Crucially, if we are interested in a state with a specific $S$ that we set as the initial $|m_{z}|$, any component after SSP that has $|m_{x}|>|m_{z}|$ does not correspond to a state with a valid $S$, and thus an ancilla readout of $|m_{x}|>|m_{z}|$ serves as a mark of spin invalidation, allowing such shots to be penalized or discarded before the costly Hamiltonian estimation.

Ancilla-assisted symmetry discrimination has been explored in several related settings. Lacroix introduced a discrete-spectra-assisted QPE construction for symmetry restoration and state preparation, illustrated through particle-number projection in a pairing model motivated by nuclear many-body physics \cite{lacroixSymmetryAssistedPreparationEntangled2020}. Siwach and Lacroix developed total-spin filtering algorithms using phase estimation of $\hat{S}^{2}$ and $\hat{S}_{z}$, together with reduced-resource alternatives \cite{siwachFilteringStatesTotal2021}. Related projection schemes employ controlled one-body rotations. Tsuchimochi et al. evaluated spin-projected molecular VQE energies by quadrature over spin-rotation kernels measured through ancilla-assisted Hadamard tests \cite{tsuchimochiSpinprojectionQuantumComputation2020}. Rule et al. used sequential single-ancilla cosine filters involving $\hat J_{z}$ and $\hat J_{x}$ for iterative $J=0$ state preparation in nuclear many-body systems \cite{ruleSimplifiedProjectionTotal2024}.

Our contribution is a spin-screening procedure for variational states with fixed $m_{z}=S_{\mathrm{tgt}}$. We use transverse-spin phase readout to identify higher-spin contaminants without direct evolution under $\hat{S}^{2}$. The screening rule applies to any allowed target spin, with rejection probabilities determined by the target and contaminating spin sectors. We use this incomplete discrimination directly in variational optimization, through which (1) contaminated trials can be rejected before further Hamiltonian measurements, or (2) higher-spin contributions can be penalized in the energy objective. These formulations optimize the original trial state without minimizing a normalized spin-projected energy. We implement both approaches within VQD for sequential excited-state calculations. 

\subsection{Spin state analysis with NISQ device}
\label{sec:calibrated_reconstruction}

The screening procedure described above will work as intended on an ideal quantum device, but it may not work as desired on noisy hardware. The rejecting or penalizing scheme assumes that the ancilla readout accurately resolves the spin projection $m_{x}$. Hardware noise can therefore cause target-spin states to be screened incorrectly. As a matter of fact, presently available hardware is quite prone to noise, and after some experiments we indeed found this issue to be significant. However, we also found that the full ancilla distribution may still retain useful information about the spin composition. More specifically, we found that noisy ancilla distributions from states in the same $(S,m_{z})$ sector retained some similar patterns when the circuit structure, depth, and hardware conditions were matched. We therefore tested the idea of using distributions measured from well-defined reference states to estimate the spin composition of any arbitrary state. We further discuss how this ``calibrated'' spin analysis may support our screening idea even with actual noisy hardware.

To see why this is possible, let us first consider an ideal distribution. For a fixed $m_{z}$, a spin sector with $S$ produces an ancilla profile determined by the probabilities $\left|d_{m_{x},m_{z}}^{S}\right|^{2}$ given by Eq. \ref{eq:wigner_expansion}:

\begin{equation}
    \mathbf{p}_{S, m_{z}}^{\mathrm{ideal}} = \begin{pmatrix}
        \vdots \\ \left|d_{-1,m_{z}}^{S}\right|^{2} \\ \left|d_{0,m_{z}}^{S}\right|^{2} \\ \left|d_{1,m_{z}}^{S}\right|^{2} \\ \vdots
    \end{pmatrix} .
    \label{eq:pure_ss_profile}
\end{equation}
We will refer to this as the pure spin-sector profile. Because a spin measurement does not couple different eigenspaces of $\hat{S}^{2}$, the ancilla profile measured from ideal hardware for a state containing multiple spin sector components can be written as
\begin{equation}
    \mathbf{p}_{\mathrm{target}}^{\mathrm{ideal}} =
    \sum_{S=0}^{S_\text{max}} w_S \mathbf{p}_{S, m_{z}}^{\mathrm{ideal}}
\end{equation}
with $w_S$ denoting the weight of each spin sector $S$, satisfying $w_S\geq0$ and $\sum_S w_S=1$.
When the pure spin-sector profiles in Eq. \ref{eq:pure_ss_profile} are compacted as columns of $\mathbf{B}_{m_{z}}^{\mathrm{ideal}}$, and when the values of $w_S$ are collected as the components a vector $\mathbf w$, the above ideal profile can be rewritten as
\begin{equation}
    \mathbf{p}_{\mathrm{target}}^{\mathrm{ideal}} = \mathbf{B}_{m_{z}}^{\mathrm{ideal}}\cdot\mathbf{w} .
    \label{eq:w_ideal}
\end{equation}
Thus, when $\mathbf{B}_{m_{z}}^{\mathrm{ideal}}$ has a full column rank, the spin-sector populations $\mathbf w$ can be uniquely determined from $\mathbf{p}_{\mathrm{target}}^{\mathrm{ideal}}$ by solving this linear system.

The same logic can also be applied even to real hardware.
For any given ansatz, we can easily prepare a spin sector with chosen values of $(S,m_z)$, and then execute the circuit operations to collect the ancilla distributions as $\mathbf{p}^{\mathrm{real}}_{S,m_{z}}$. After repeating this for different values of $S$, the results can be compacted as $\mathbf{B}^{\mathrm{real}}_{m_{z}}$. With this preparation, when there is an unknown state defined by different parameters with the same ansatz, it may be processed to produce an ancilla distribution $\mathbf{p}^{\mathrm{obs}}_{r}$.
Then, its spin-sector populations $\mathbf{w}^\mathrm{unknown}$ can be deduced by
\begin{equation}
\mathbf{p}^{\mathrm{obs}}_{r} \simeq
\mathbf{B}^{\mathrm{real}}_{m_{z}}\cdot\mathbf{w}^\mathrm{unknown} .
\end{equation}
Operationally, spin-sector populations can be reconstructed by nonnegative least squares:
\begin{equation}
\hat{\mathbf{w}}=\underset{\mathbf{w}\geq 0}{\operatorname{argmin}}\left\| \mathbf{B}^{\mathrm{real}}_{m_{z}}\cdot\mathbf{w}-\mathbf{p}^{\mathrm{obs}}_{r}\right\|^2,
\label{eq:w-hat}
\end{equation}
with the constraint of $\sum_S\hat{w}_S=1$.

Of course, the statistics of $m_{x}$ can also be measured by direct occupation measurements after the same rotation of the spin basis. However, doing so will collapse the state with respect to $\hat S_z$. In contrast, our scheme performs post-selection through only ancilla measurements and can preserve the system description inherited from $|\psi_0\rangle$ in Figure \ref{fig:R_{x}_screening} with respect to its energy for subsequent measurements. In fact, its philosophy is similar to that of the particle number post-selection.\cite{lacroixSymmetryAssistedPreparationEntangled2020} With ideal hardware, our formalism guarantees that any $S \le S_\text{tgt}$ will always be accepted. In the case of $S > S_\text{tgt}$ that falsely pass the filtering, a circuit with repeated ancilla measurements can be constructed to reduce the pass probability. Under noisy measurements, we can build a decision rule based on calibrated measurement statistics to decide whether to reject the current parameter set or continue its evaluation. The rule will require sufficient evidence that the off-target spin weight exceeds a chosen tolerance, with the aim of preventing falsely rejecting $S \le S_\text{tgt}$.

The spirit of this procedure is calibration. Namely, $\mathbf{B}^{\mathrm{real}}_{m_{z}}$ is prepared beforehand with well-defined states and then an unknown state is analyzed based on it. Quite naturally, as in other calibrations, $\mathbf{B}^{\mathrm{real}}_{m_{z}}$  will be specific to the circuit family, ansatz layer, qubit layout, and even hardware condition and therefore should not be assumed to be universally transferable. In our experiments that we will discuss in a later part, the population analyses with calibration preparation for $\mathbf{B}^{\mathrm{real}}_{m_{z}}$ and actualy execution for $\mathbf{p}^{\mathrm{obs}}_{r}$ were performed in pairs to minimize the influence of temporal device drift. The circuit construction and hardware mapping are detailed in Figure S10.

\subsection{Computational protocol}
We first tested spin-selective VQD using statevector simulations. We also used noisy simulations and hardware measurements to evaluate the reconstruction of spin-sector populations from calibrated ancilla distributions. These noisy benchmarks evaluated the sensing and reconstruction steps separately from the full VQD optimization.

First, to demonstrate the proposed spin-preserving protocol with VQD, we applied it to small benchmark molecules, \ce{LiH} and \ce{BeH2} \cite{omalleyScalableQuantumSimulation2016, roosCompleteActiveSpace1980}. The molecular geometries were parameterized as $\mathbf{R}(\lambda) = \mathbf{R}_{0} + \lambda\cdot \Delta\mathbf{R}$. For \ce{LiH}, the equilibrium bond length was set to 1.360~\AA, with $\Delta \mathbf{R}$ displacing the distance of \ce{Li} and \ce{H} atoms along the molecular axis by 1.000~\AA, over $\lambda\in[-1.0, 1.0]$. For \ce{BeH2}, $\mathbf{R}_{0}$ corresponds to the equilibrium geometry of linear \ce{BeH2} with the \ce{Be}--\ce{H} bond length of 1.334 \AA. For the symmetric stretch, $\Delta\mathbf{R}$ was set to displace the two hydrogen atoms in opposite directions along the molecular axis over $\lambda \in [-1.0, 1.0]$ with $\Delta \mathbf{R}$ of 0.700 \AA. For the antisymmetric stretch, the two atoms were displaced in the same direction over $\lambda \in [0.0, 1.0]$ with the same $\Delta \mathbf{R}$. These ranges of geometries were adopted as standard test cases in previous studies with quantum algorithms for electronic structures \cite{kandalaHardwareefficientVariationalQuantum2017, omalleyScalableQuantumSimulation2016, mccleanTheoryVariationalHybrid2016}. All calculations in this work employed the STO-3G basis set \cite{hehreSelfConsistentMolecularOrbitalMethods1969}. For each molecule, we defined an active space that can capture the chemically relevant valence orbitals and froze the remaining core and non-bonding or weakly participating $\pi$ valence orbitals at the mean-field level. The resulting active-space Hamiltonians contained 3 spatial orbitals and 2 active electrons for \ce{LiH}, and 4 spatial orbitals and 4 active electrons for \ce{BeH2}.

As classical references, we computed state-specific complete active space configuration interaction (CASCI) energies \cite{knowlesNewDeterminantbasedFull1984} with exactly the same active space definitions to assess the performance of our approach. On the quantum algorithm side, we employed the parametrized circuit shown in Figure~\ref{fig:R_{x}_screening} with 3--10 layers of the SSP ansatz circuit. The parameters were initialized from a zero-mean normal distribution with a standard deviation of 0.3. 
For completeness, Algorithm~\ref{alg:spin_selective_measurement_shot_simulator} summarizes a strict shot-based rejection protocol for an error-free quantum device. For each trial state, the ancilla register is sampled before the Hamiltonian is measured. If any sampled outcome satisfies $|m_{x}|>S_{\mathrm{tgt}}$, the current trial is rejected and assigned the prescribed penalty; otherwise, the Hamiltonian and deflation terms are evaluated. This protocol treats the ancilla outcome as a hard spin-invalidation witness and therefore represents the most direct measurement-based implementation of the proposed screening rule. For each numerical experiment, the variational optimization was independently repeated ten times, and the run yielding the lowest value of the corresponding objective function was retained.

\begin{algorithm}
\caption{Spin-filtering VQD with strict shot-based rejection on an error-free quantum device.}
\label{alg:spin_selective_measurement_shot_simulator}
\begin{algorithmic}[1]
\Require Hamiltonian $\hat{H}=\sum_{P\in\mathcal{P}} c_{P} P$, target spin quantum number $S_{\mathrm{tgt}}$, order of desired excited state $k$, previous excited statevectors $|\psi^{(0)}\rangle,\cdots|\psi^{(k-1)}\rangle$, number of shots per Pauli string $N_{\mathrm{shot}}$, parameters $\Theta$ initialized to Gaussian random numbers
\Ensure $k$-th excited statevector of target spin quantum number $S_{\mathrm{tgt}}$, $|\psi^{(k)}_{l_{s}}(\Theta)\rangle$ \Repeat
    \State $|\psi(\Theta)\rangle \gets \mathbf{SSP}(\Theta)|\psi_{0}\rangle$
    \State $\begin{aligned}[t]
    |\Psi(\Theta)\rangle \gets{}
    \bigl(I\otimes\mathcal{QFT}_{m_x}^{\dagger}\bigr)\, &\mathcal{QPE}(\hat S_x)\\
    &\quad\quad\cdot\;\bigl(|\psi(\Theta)\rangle\otimes|0\rangle_{m_x}\bigr)
    \end{aligned}$
    \State $\mathcal{L}\gets0$
    \ForAll{$P \in \mathcal{P}$}
        \For{$i=1,\cdots,N_{\mathrm{shot}}$}
            \State $m_{x,\mathtt{iter}} \gets$ Measure $m_{x}$ register of $|\Psi(\Theta)\rangle$ 
            \If{$|m_{x,\mathtt{iter}}| > S_{tgt}$}
                \State $\mathcal{L}\gets\|\hat{H}\|$
                \State \textbf{break}
            \Else
                \State $l\gets\langle\Psi(\Theta)|P\otimes I_{m_{x}}|\Psi(\Theta)\rangle$
                \State $\mathcal{L}\gets\mathcal{L}+c_{P}\cdot{l}/{N_{\mathrm{shot}}}$
            \EndIf
        \EndFor
        \If{$i \leq N_{\mathrm{shot}}$}
            \State \textbf{break}
        \EndIf
    \EndFor
    \For{$j=0,\cdots,k-1$}
        \State $l\gets\left\langle\psi\left(\Theta\right)\middle|\psi^{(j)}\right\rangle\!\left\langle\psi^{(j)}\middle|\psi\left(\Theta\right)\right\rangle$
        \State $\mathcal{L}\gets\mathcal{L}+c_{j}\,\cdot l$
    \EndFor    
    \State Update $\Theta$
\Until{$\mathcal{L}$ is minimized}
\end{algorithmic}
\end{algorithm}
However, when simulating in a statevector simulator, the halting scheme in Algorithm  \ref{alg:spin_selective_measurement_shot_simulator} (Line 10) is inefficient. This can be improved by deterministic penalty formulation scheme with Algorithm  \ref{alg:spin_selective_measurement_statevector}, with which the hard-rejection rule can be replaced by a deterministic expectation-value objective constructed from the extended Hamiltonian $\hat{H}_{\mathrm{ext}}$,
\begin{equation}
\hat{H}_{\mathrm{ext}}
= \left(\hat{H}-\|{\hat{H}}\|\cdot I\right)\otimes \hat{D}_\mathrm{anc} \, + \|\hat{H}\|\cdot I \otimes I_\mathrm{anc} ,
\label{eq:H_ext}
\end{equation}
with diagonal operator $D_\mathrm{anc}$ acting on the ancilla register,
\begin{equation}
\hat{D}_\mathrm{anc} = \sum_{j=-2^{n_\mathrm{anc}-1}}^{2^{n_\mathrm{anc}-1}-1} d(j)\,\ket{j}\bra{j},
\label{eq:D_anc}
\end{equation}
together with the penalty weights
\begin{equation}
d(j)=
\begin{cases}
1, & \text{if} \;\; |m_{x}(j)|\leq |m_{z}| \\
c_{\mathrm{penalty}}, & \text{otherwise}
\end{cases}
\label{eq:penalty_weight}
\end{equation}
and $I_\mathrm{anc}$ representing the identity in the ancilla space.
The construction is designed such that $\hat{H}-\|\hat{H}\|\cdot I$ is negative semi-definite, which ensures that adopting $D_\mathrm{anc}$ with a penalty weight $c_{\text{penalty}}\in(-\infty, 1)$ shifts the effective energy of spin-invalid states upward relative to valid ones. For this purpose, the spectral norm $\|\hat{H}\|$ is not exactly needed, but any large enough number will suffice to guarantee the negative semi-definiteness. If we imagine $\mathcal{P}$ as a set of Pauli strings that are needed in building $\hat H$, namely if we write the Hamiltonian as $\hat{H}=\sum_{P\in\mathcal{P}}c_{P}P$, because $\|P\|=1$ for any Pauli operator, the following inequality is always guaranteed:
\begin{equation}
    \|\hat{H}\| \leq \sum_{P\in\mathcal{P}}|c_{P}| .
\end{equation}
Thus, $c_{H} \equiv \sum_{P} |c_{P}|$ can be used in the place of $\| \hat {H} \|$ in Eq.\ \ref{eq:H_ext}. This value is directly available from the Pauli decomposition without additional diagonalization cost.
To keep the penalty scale comparable to the one used in the VQD overlap terms and to avoid introducing an additional tuning parameter, we simply fixed it as $c_{\mathrm{penalty}}=0$ throughout this work. For the VQD overlap penalty, we employed the convention of using absolute value of the ground-state energy, $|E_0|$ \cite{higgottVariationalQuantumComputation2019}.

\begin{algorithm}[t]
\caption{Spin-filtering VQD with statevector-based simulation.}
\label{alg:spin_selective_measurement_statevector}
\begin{algorithmic}[1]
\Require Extended Hamiltonian $\hat{H}_\mathrm{ext}$, target spin quantum number $S_\mathrm{tgt}$, index of desired excited state $k$, previous excited statevectors $|\psi^{(0)}\rangle,\dots,|\psi^{(k-1)}\rangle$, parameters $\Theta$  initialized to Gaussian random numbers
\Ensure $k$-th excited statevector of target spin quantum number $S_\mathrm{tgt}$, $|\psi^{(k)}_{l_{s}}(\Theta)\rangle$
\Repeat
    \State $|\psi(\Theta)\rangle \gets \mathbf{SSP}(\Theta)|\psi_{0}\rangle$
    \State $\begin{aligned}[t]
    |\Psi(\Theta)\rangle \gets{}
    \bigl(I\otimes\mathcal{QFT}_{m_x}^{\dagger}\bigr)\, &\mathcal{QPE}(\hat S_x)\\
    &\quad\quad\cdot\;\bigl(|\psi(\Theta)\rangle\otimes|0\rangle_{m_x}\bigr)
    \end{aligned}$
    \State $\mathcal{L}\gets\langle\Psi(\Theta)|\hat{H}_\mathrm{ext}|\Psi(\Theta)\rangle $
    \For{$j=0,\cdots,k-1$}
        \State $l\gets\left\langle\psi\left(\Theta\right)\middle|\psi^{(j)}\right\rangle\!\left\langle\psi^{(j)}\middle|\psi\left(\Theta\right)\right\rangle$
        \State $\mathcal{L}\gets\mathcal{L}+c_{j}\,\cdot l$
    \EndFor    
    \State Update $\Theta$
\Until{$\mathcal{L}$ is minimized}
\end{algorithmic}
\end{algorithm}

Next, we evaluated the calibration-assisted reconstruction described in Sec.~\ref{sec:calibrated_reconstruction} using noisy shot simulations and hardware measurements. These benchmarks retained only the ancilla readout following state preparation and sensing, omitting Hamiltonian measurements to assess reconstruction accuracy independently of Hamiltonian-sampling errors. Algorithm~\ref{alg:calibrated_spin_reconstruction} summarizes the benchmark procedure.

\begin{algorithm}[t]
\caption{Reconstruction of spin-sector population by calibration.}
\label{alg:calibrated_spin_reconstruction}
\small
\begin{algorithmic}[1]

\Require Reference states $\{|\psi_{S,m_{z}}\rangle\}$, target state $|\psi_{\mathrm{tar}}\rangle$ with fixed $m_{z}$, number of shots $N_{\mathrm{shot}}$

\Ensure Reconstructed spin-sector population of target state $\hat{\mathbf{w}}_\mathrm{tar}$

    \For{$S=|m_z|,\cdots,S_{\max}$}
        \State $|\Psi_{S,m_{z}}\rangle \gets \left\{ I\otimes\mathcal{QFT}_{m_{x}}^{\dagger} \cdot\mathcal{QPE}(\hat S_{x}) \right\} |\psi_{S,m_{z}}\rangle \otimes|0\rangle_{m_{x}}$
        \State $\mathbf{n}_{S,m_z} \gets$ Measure $m_{x}$ register of $|\Psi_{S,m_{z}}\rangle$ with $N_{\mathrm{shot}}$ shots
        \State $\left[\mathbf{p}_{S,m_{z}}^{\mathrm{cal}}\right]_{m_{x}} \gets n_{S,m_{z}}(m_{x})/N_{\mathrm{shot}}$
    \EndFor

    \State $\mathbf{B}_{m_z}^{\mathrm{cal}} \gets \left[ \mathbf{p}_{|m_z|,m_z}^{\mathrm{cal}} \ \cdots\ \mathbf{p}_{S_{\max},m_z}^{\mathrm{cal}} \right]$

    \State $|\Psi_\mathrm{tar}\rangle \gets \left\{ I\otimes\mathcal{QFT}_{m_x}^{\dagger} \cdot\mathcal{QPE}(\hat S_x) \right\} |\psi_\mathrm{tar}\rangle\otimes|0\rangle_{m_x}$
    \State $\mathbf{n}_\mathrm{tar} \gets$ Measure $m_{x}$ register of $|\Psi_\mathrm{tar}\rangle$ with $N_\mathrm{shot}$ shots
    \State $\mathbf{p}_\mathrm{tar}^{\mathrm{obs}} \gets
        \mathbf{n}_\mathrm{tar}/N_{\mathrm{shot}}$
    \State $\tilde{\mathbf{w}}_\mathrm{tar} \gets
        \underset{\mathbf{w}\geq 0}{\operatorname{argmin}}
        \left\| \mathbf{B}_{m_z}^{\mathrm{cal}}\mathbf{w}
        - \mathbf{p}_\mathrm{tar}^{\mathrm{obs}} \right\|^2$
    \State $\hat{\mathbf{w}}_\mathrm{tar} \gets
        \tilde{\mathbf{w}}_\mathrm{tar} \big/
        \displaystyle\sum_{S=|m_z|}^{S_{\max}}\tilde{w}_{\mathrm{tar},\,S}$

\end{algorithmic}
\end{algorithm}

To restore the Hamiltonian-measurement step and close the calibrated VQD loop with noisy hardware, we need a scheme to put penalties to a running parameter set. In the case of ideal hardware, $\hat D_{\mathrm{anc}}$ and $d(j)$ in Eqs. \ref{eq:D_anc} and \ref{eq:penalty_weight} defined with pre-set penalty weights play such a role. We can also adopt
$    [\mathbf{d}^{\mathrm{ideal}} (\mathbf{B}_{m_{z}}^{\mathrm{ideal}} ) ]_S
$,
with $\mathbf{d}^{\mathrm{ideal}}$ denoting a row vector formed from $d(j)$ to represent penalties on a state with total angular momentum $S$. Thus, on a noisy hardware, we propose to obtain calibrated ancilla weights by solving minimization:
\begin{equation}
    \mathbf{d}^{\mathrm{cal}} = \underset{\mathbf {d}}{\operatorname{argmin}} \left\| \mathbf{d}\, \mathbf{B}_{m_{z}}^{\mathrm{real}} - \mathbf{d}^{\mathrm{ideal}} \mathbf{B}_{m_{z}}^{\mathrm{ideal}} \right\|^{2}.
\end{equation}
When $\mathbf{B}_{m_{z}}^{\mathrm{real}}$ has full column rank, the solution simplifies to
\begin{equation}
    \mathbf{d}^{\mathrm{cal}}
    = \mathbf{d}^{\mathrm{ideal}} \,\mathbf{B}_{m_z}^{\mathrm{ideal}} \,\left(\mathbf{B}_{m_z}^{\mathrm{real}}\right)^{+},
    \label{eq:d^cal}
\end{equation}

where $+$ denotes the Moore--Penrose pseudoinverse. These weights define the ancilla operator calibrated in a noisy hardware:
\begin{equation}
    \hat{D}_{\mathrm{anc}}^{\mathrm{cal}} = \sum_{j} d_{j}^{\mathrm{cal}} |j\rangle\langle j|.
\end{equation}

Substituting the above result for the ideal ancilla operator in Eq. \ref{eq:H_ext} gives a new extended Hamiltonian
\begin{equation}
    \hat{H}_{\mathrm{ext}}^{\mathrm{cal}} = \left(\hat{H}-\|\hat{H}\|\cdot I\right)\otimes\hat{D}_{\mathrm{anc}}^{\mathrm{cal}}+\|\hat{H}\|\cdot I\otimes I_{\mathrm{anc}} .
\end{equation}
Of course, $\hat H_{\mathrm{ext}}^{\mathrm{cal}}$ will have different forms for different HEA layer numbers.
In a shot-based evaluation, the ancilla register and a Hamiltonian Pauli term $P$ are measured in the same circuit execution. if $j_{i}$ is the ancilla outcome and $p_{i}\in\{-1,1\}$ is the corresponding Pauli outcome of shot $i$, their jointly weighted estimator is
\begin{equation}
    \left\langle P \otimes \hat{D}_{\mathrm{anc}}^{\mathrm{cal}}\right\rangle = \frac{1}{N_{\mathrm{shot}}}\sum_{i=1}^{N_{\mathrm{shot}}}{d_{j_{i}}^{\mathrm{cal}}p_{i}}
\end{equation}
These estimates provide the Hamiltonian contribution to the calibrated extended Hamiltonian objective used in Algorithm \ref{alg:spin_selective_measurement_statevector}. 

\section{Results and Discussion}

\subsection{Energetic behavior and spin selectivity}

\begin{figure}
    \centering
    \includegraphics[width=0.5\linewidth]{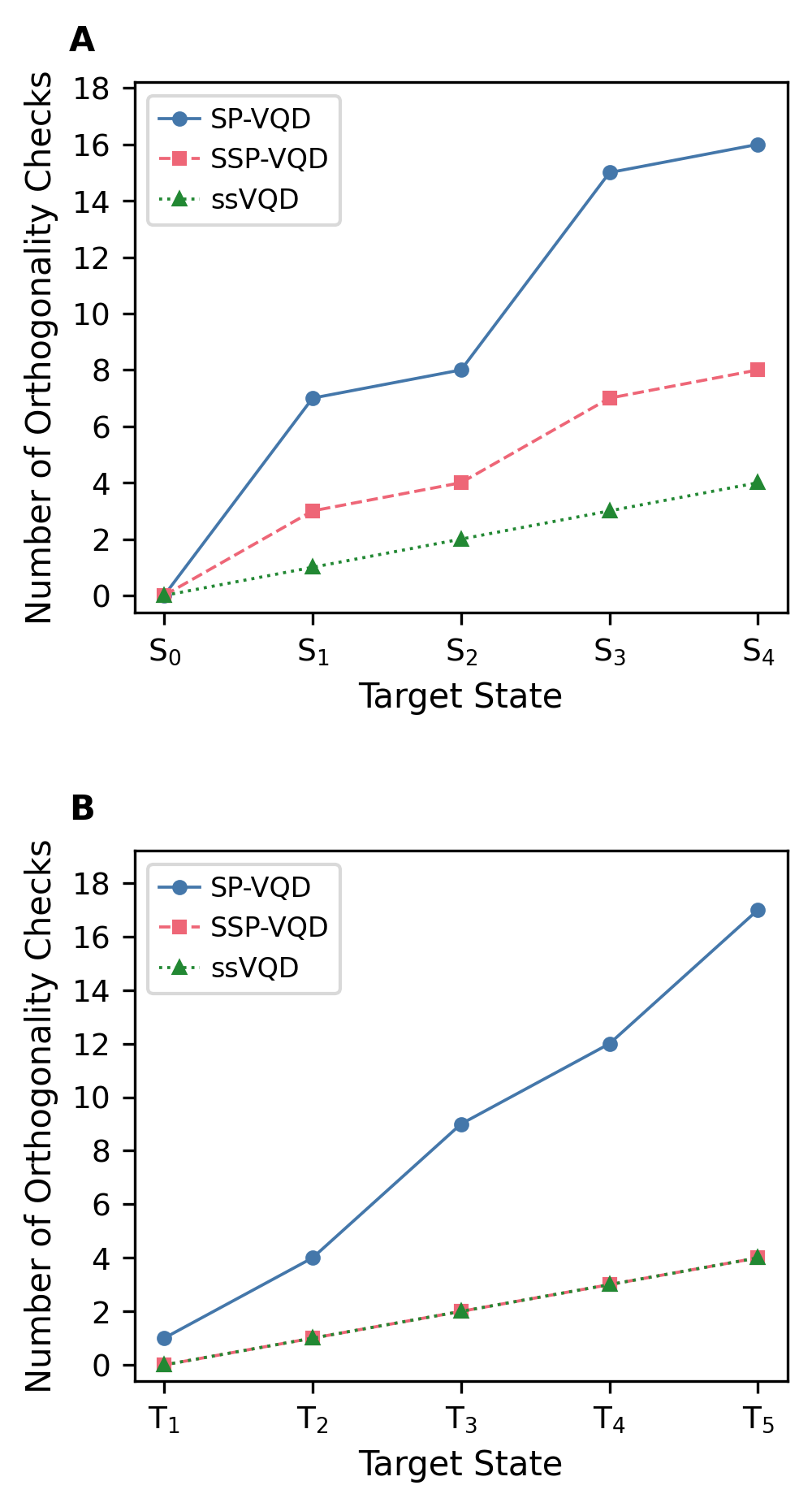}
    \caption{Number of overlap terms required in the VQD deflation penalty for each target excited state in the singlet (A) and triplet (B) manifolds of \ce{BeH2}.}
    \label{fig:orth_check_numbers}
\end{figure}

To evaluate the performance of our proposed method, we compared three approaches. The first is conventional VQD using the SP ansatz (VQD/SP), and the second uses the SSP ansatz (VQD/SSP), which restricts the variational search space to a fixed $m_{z}$ sector. Our proposed method further suppresses unwanted spin components through spin filtering, and is therefore referred to as spin-filtering VQD with the SSP ansatz (sfVQD/SSP).

We first examine whether our approach successfully reduces the number of orthogonality checks required in VQD by restricting the number of states that need to be considered during deflation. As shown in Figure~\ref{fig:orth_check_numbers}, sfVQD/SSP requires approximately four times fewer deflation overlap terms than VQD/SP and two times fewer than VQD/SSP for the singlet states of \ce{BeH2}. This reduction arises from the progressively stronger symmetry restrictions, which reduce the number of unwanted spin states that need to be considered during deflation. Reducing these overlap terms can lower the associated circuit overhead and may also limit the propagation of numerical errors from previously converged states. Moreover, spin filtering allows states with different spin multiplicities to be distinguished even when their energies are close or degenerate.

\begin{figure}
    \centering
    \includegraphics[width=1.0\linewidth]{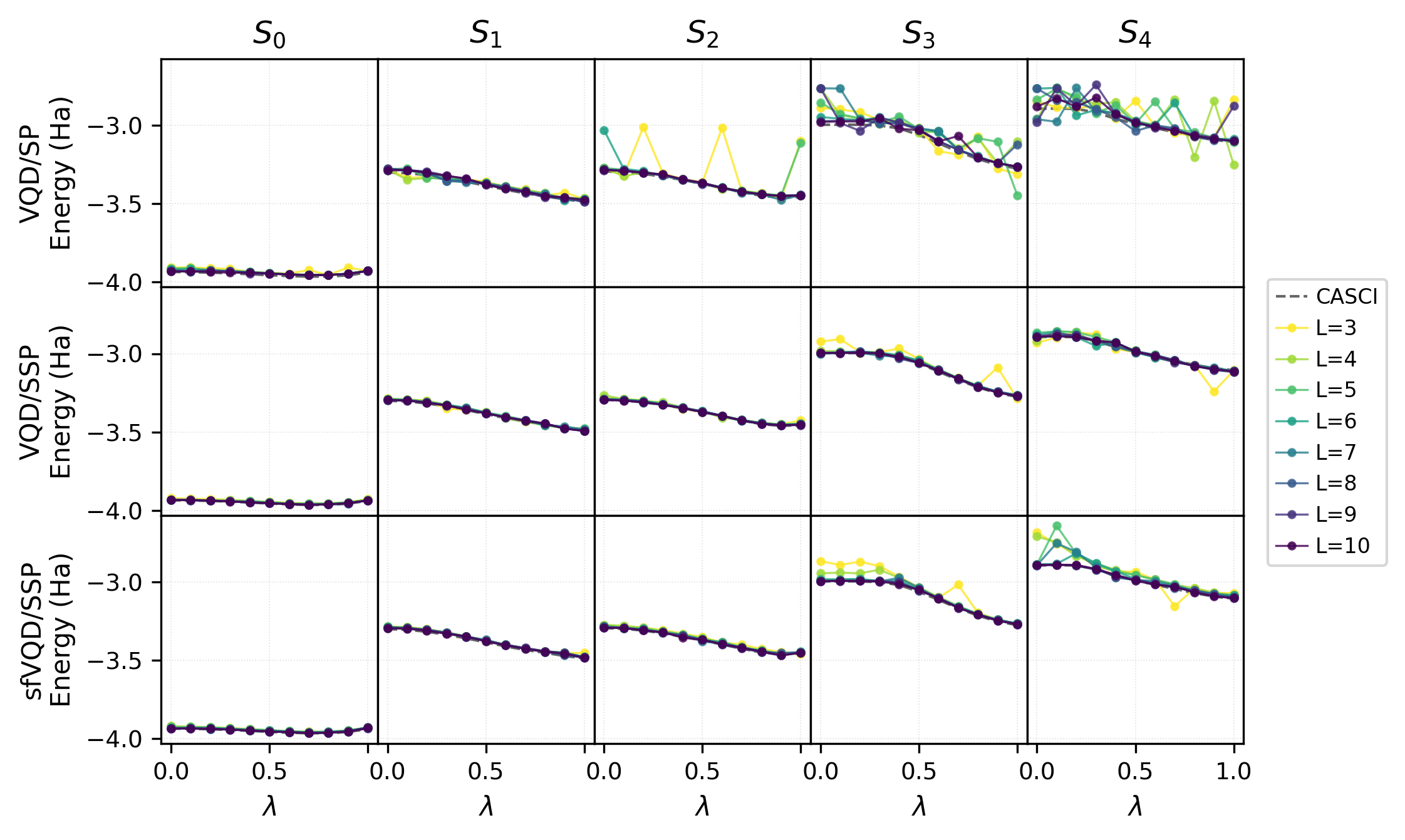}
    \caption{Potential energy surfaces of $\text{S}_0$ -- $\text{S}_4$ states for antisymmetrically stretched \ce{BeH2} obtained with VQD/SP (top), VQD/SSP (middle), and sfVQD/SSP (bottom) in comparison with CASCI energies. Color-coded $L$ represents the adopted repetition number of the ansatz circuit.}
    \label{fig:graphs_asym}
\end{figure}

We next calculated potential energy surfaces using statevector simulations. Figure~\ref{fig:graphs_asym} compares the results for the $S_{0}$–$S_{4}$ states of \ce{BeH2} along an antisymmetric stretching coordinate, obtained using the three approaches with different HEA layer counts denoted with $L$. In general, increasing the number of ansatz layers improved the stability of the calculated energy profiles, while methods requiring fewer deflation overlap terms tended to produce more consistent results. At some geometries, however, VQD/SSP produced smoother energy profiles than sfVQD/SSP. This difference diminished as the number of ansatz layers increased, suggesting that greater ansatz expressibility can mitigate the unfavorable effects of spin filtering on the optimization. Corresponding potential energy surfaces for symmetrically stretched \ce{BeH2} and \ce{LiH} are presented in SI Figures S3--S7.

\begin{figure}
    \centering
    \includegraphics[width=1.0\linewidth]{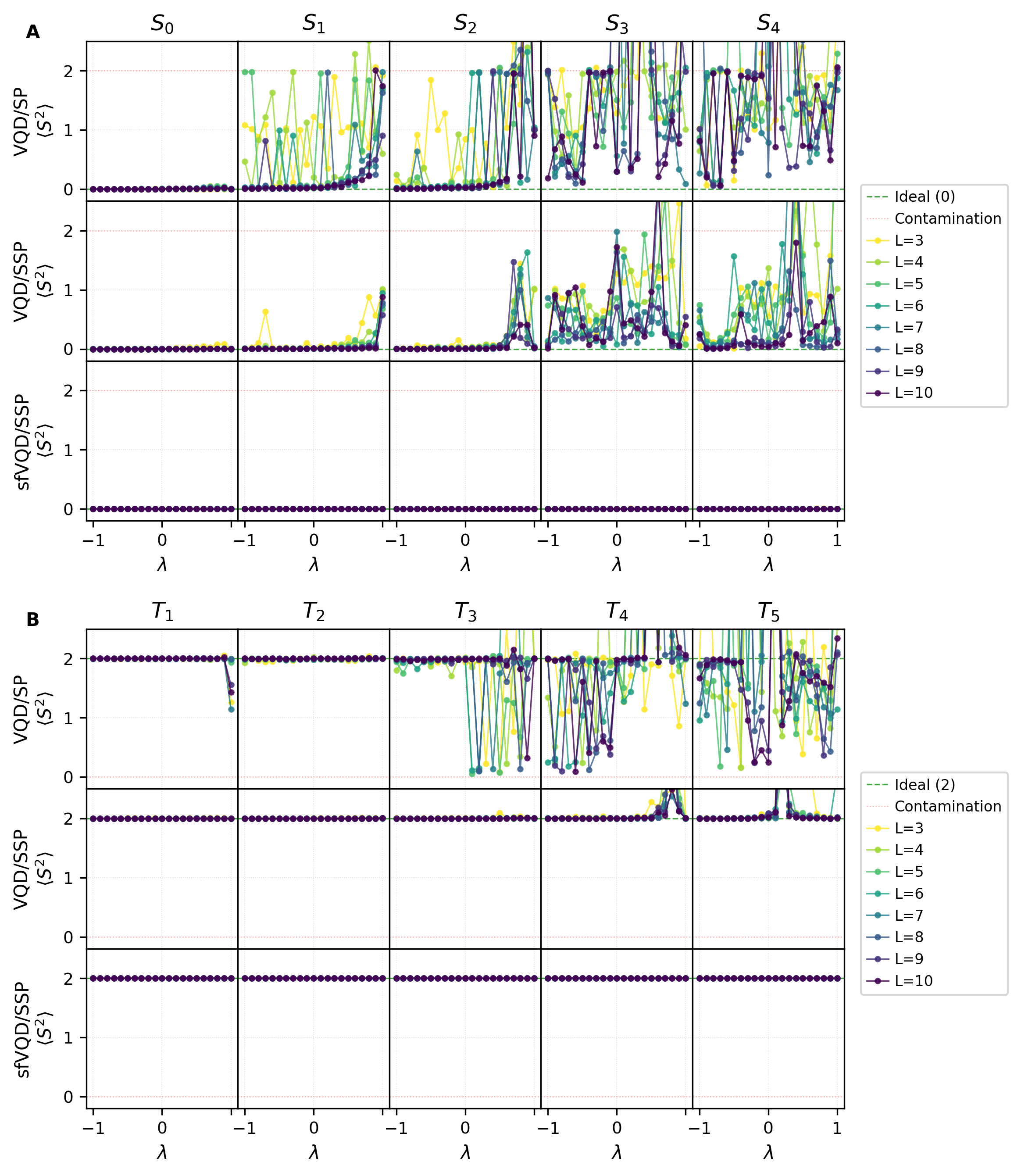}
    \caption{Spin diagnostic with $\langle \hat{S}^{2}\rangle$ for singlet (A) and triplet (B) states of symmetrically stretched \ce{BeH2}, performed on VQD/SP (top), VQD/SSP (middle), and sfVQD/SSP (bottom) results. Color-coded $L$ represents the adopted repetition number of the ansatz circuit.}
    \label{fig:graphs_sym_s2}
\end{figure}

We then examined how effectively sfVQD/SSP suppresses spin contamination. Figure~\ref{fig:graphs_sym_s2} compares the calculated $\langle \hat{S}^2 \rangle$ values obtained using the three approaches at different $L$ values. Both VQD/SP and VQD/SSP exhibited spin contamination, in some cases starting from the first excited state, and increasing the number of ansatz layers alone did not consistently eliminate these deviations. In contrast, sfVQD/SSP maintained $\langle \hat{S}^{2} \rangle$ values close to those expected for the target spin sectors. Similar improvements in spin selectivity were observed for \ce{LiH} and antisymmetrically stretched \ce{BeH2}, as shown in SI Figures S8 and S9.

This improvement arises from the additional $\hat{S}_{x}$-based screening step, which probabilistically identifies contributions from unwanted higher-spin sectors. The screening procedure does not act as an exact spin projector, but instead biases the variational optimization toward the desired spin sector. As a result, sfVQD/SSP can obtain excited states with more reliable spin character even when VQD/SSP alone remains susceptible to spin contamination. This approach is also compatible with the shot-based measurement framework of NISQ algorithms.

So far, our statevector simulations have demonstrated that spin filtering can reduce the deflation overhead while maintaining the desired spin character of the calculated excited states. However, on real quantum hardware, noise can distort the spin information obtained from measurements. As our screening procedure relies on ancilla readout, it is important to determine whether this information remains accessible under hardware noise. In the following section, we examine the measured spin distributions on a quantum processor and investigate whether spin-sector populations can be reconstructed through our calibration scheme.

\subsection{Calibration-based reconstruction on noisy hardware}
\label{sec:response_transferability}

We next tested whether calibration could properly recover spin-sector populations even from noisy hardware measurements. Recall that spin-sector populations here refer to the decomposition into total-spin $S$ sectors of the state prepared by an ansatz circuit with a specific parameter set obtained for a given $m_z$. For the tests, we focused on the spin readout and reconstruction stages of our screening procedure. To assess the reconstruction accuracy, we compared the reconstructed populations with those calculated from the corresponding ideal target statevectors.

On noisy hardware, we found that the additional entangling operations required for direct $\hat{S}_{x}$ readout quickly degraded the measured spin profiles. We therefore used occupation counting to reduce the additional entangling operations on the system register, where the variational state is stored. We first rotated the system into the $x$-spin basis, whose spin states are $|\pm\rangle=(|\alpha\rangle\pm|\beta\rangle)/\sqrt{2}$. We then used separate ancilla registers to count their occupations, $\hat{n}_{+}$ and $\hat{n}_{-}$. These occupations give both the transverse-spin projection and the total particle number through the relations

\begin{eqnarray}
\hat S_{x} &=& \frac{\hat n_{+}-\hat n_{-}}{2}, \\
\hat N &=& \hat n_{+}+\hat n_{-} .
\end{eqnarray}
Thus, in the ideal limit, the joint occupation distribution contains the same transverse-spin information as the direct $\hat{S}_{x}$ measurement, together with a particle-number diagnostic. The four-electron, four-orbital active space of \ce{BeH2} was represented by eight system qubits, with two three-qubit counting registers resolving occupations from zero to four without aliasing in the Fourier transform. The circuits were executed on the IBM Nighthawk processor \texttt{ibm\_miami}, using 1024 shots per circuit and SSP layer counts $L=3,5,7,$ and $10$.
For each $L$, the calibration references and target circuits were submitted together in a paired bundle with a common initial qubit layout. Ten target parameter sets were examined for both $m_{z}=0$ and $m_{z}=1$ cases. Their spin-sector populations $\hat{\mathbf{w}}$ were reconstructed using the nonnegative least-squares procedure described above in Eq. \ref{eq:w-hat}. The calibration basis contained singlet, triplet, and quintet references for $m_{z}=0$, and triplet and quintet references for $m_{z}=1$. All measured outcomes were retained in the reconstruction, including those outside the expected particle-number sector.

\begin{figure}
    \centering
    \includegraphics[width=0.5\linewidth]{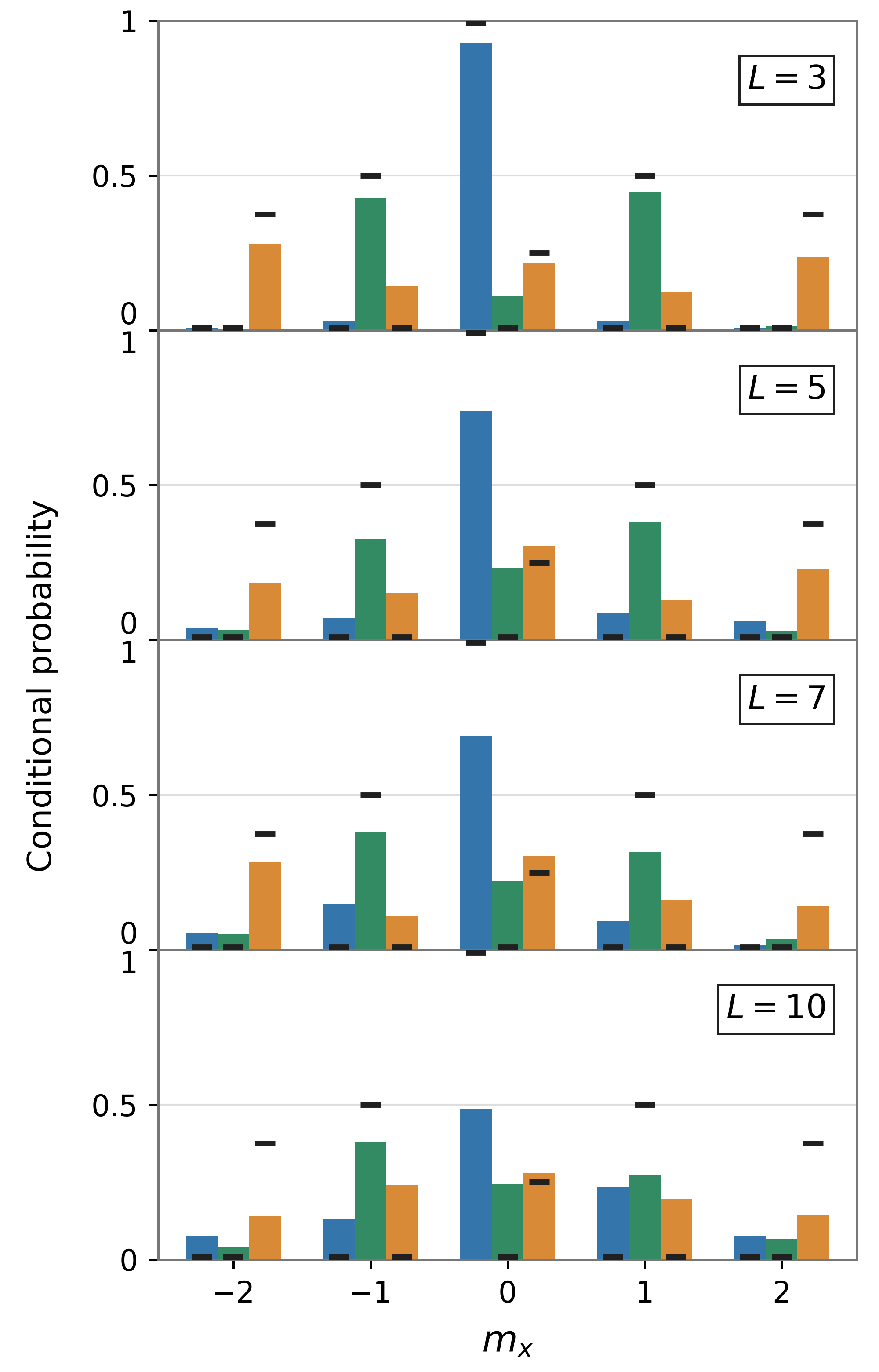}
    \caption{
    Distributions of $m_x$ obtained by calibrated measurements on singlet (blue), triplet (green), and quintet (orange) reference states generated with $m_z=0$ and $n_{+}+n_{-}=4$, compared with ideal values $|d^{S}_{m_x,m_z}|^2$ (black bars). The adopted ansatz layer count ($L$) is provided as inset in each panel.
    }
    \label{fig:hardware_spin_profiles_theory}
\end{figure}

\begin{figure}[htbp]
    \centering
    \includegraphics[width=0.65\linewidth]{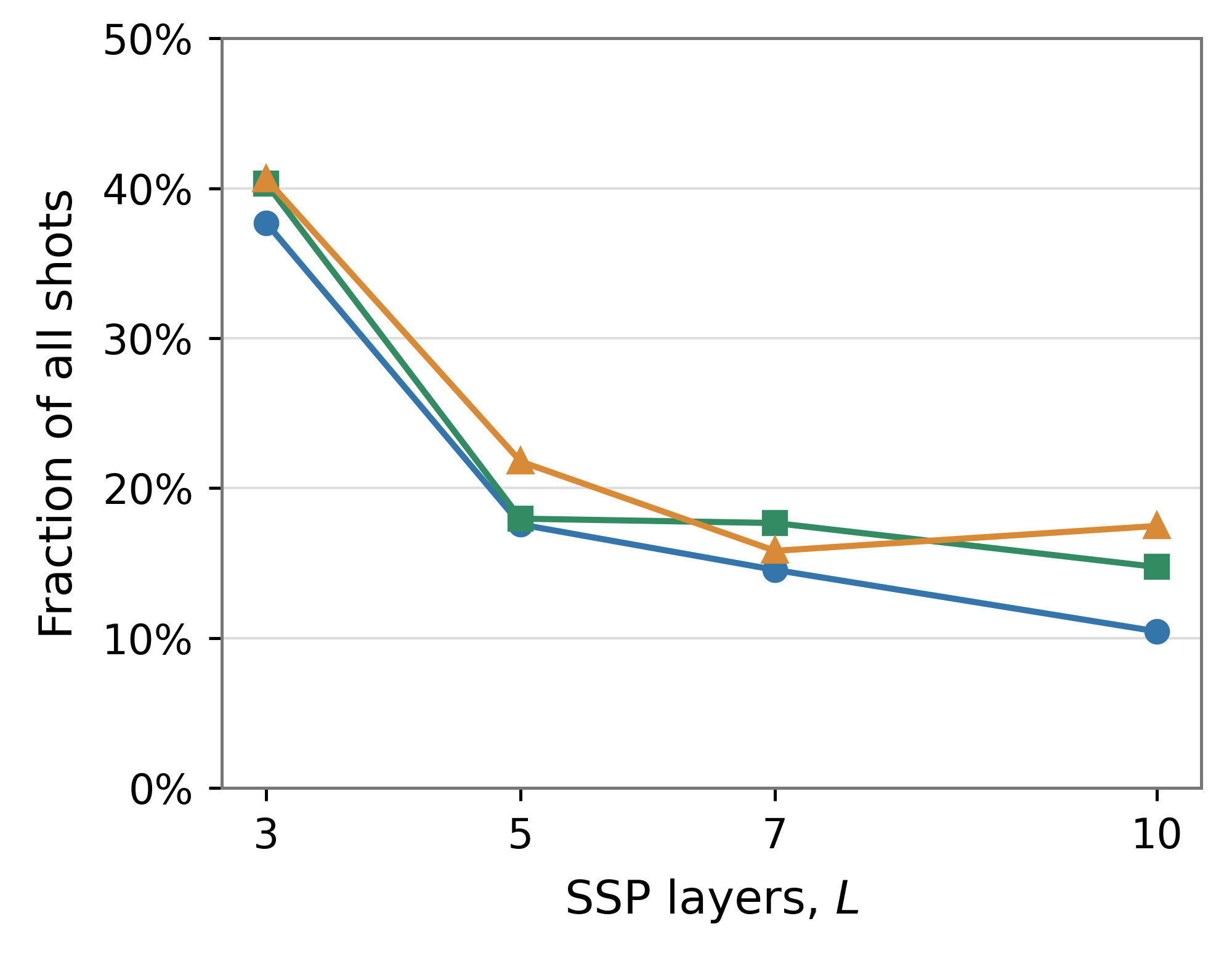}
    \caption{Fractions of measurement outcomes satisfying $n_{+}+n_{-}=4$, at different ansatz layer counts $L$ for singlet (blue), triplet (green), and quintet (orange) reference states with $m_z=0$. Each fraction is evaluated relative to all measurement shots. The count-based diagnostic included all errors in state preparation, sensing, and readout.
    }
    \label{fig:hardware_four_count_fraction}
\end{figure}

Figures~\ref{fig:hardware_spin_profiles_theory} illustrates how the measured spin profile diagnostics change with the number of ansatz layers. At $L=3$, the spin-dependent features remain distinguishable despite hardware noise, with the singlet distribution concentrated near $m_{x}=0$ and the triplet distribution retaining peaks near $m_{x}=\pm1$. As the number of layers increases, these features become less pronounced, and the measured distributions deviate further from their ideal profiles.
Meanwhile, Figure~\ref{fig:hardware_four_count_fraction} shows that the fraction of outcomes satisfying the correct particle number ($n_{+}+n_{-} = 4$) decreases with increasing circuit depth. For the three reference spin-sectors for $m_{z}=0$, this fraction at $L=3$ is already as low as 37\%--41\% and further deteriorates down to $\sim$10\% at $L=10$. From this, we can infer that the hardware noise was already immense at $L=3$, but our filtering scheme adapted to the situation quite well with reasonable spin-state distinguishability even up to $L=7$.

\begin{figure}
\centering
\includegraphics[width=0.5\linewidth]{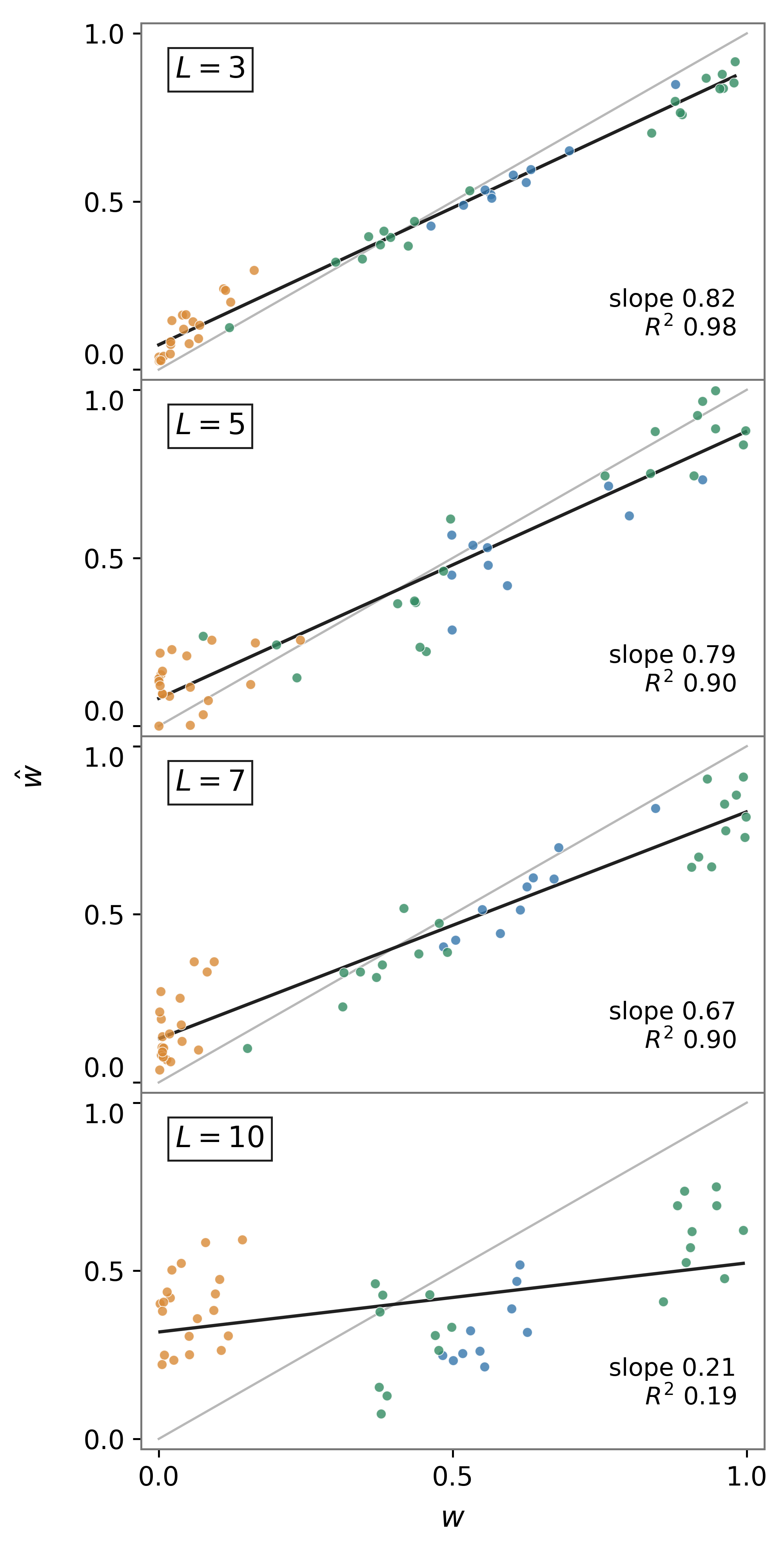}
\caption{Spin-sector populations reconstructed from paired hardware calibration versus ideal statevector populations. Each panel shows ten test results for $m_z = 0$ and additional ten for $m_z = 1$ with other circuit parameters set to random. In each panel, a test with $m_z=0$ generated three dots with $S=0,1,2$, while a test with $m_z=1$ generated two dots with $S=1,2$. Colors indicate $S=0$ (blue), $S=1$ (green), and $S=2$ (orange). Black lines show linear regressions over the data, while gray lines show diagonal perfect correlation lines. The slopes and the $R^2$ values of the correlations are also displayed. See Eqs. \ref{eq:w_ideal} and \ref{eq:w-hat} for the definitions of $\mathbf{w}$ and $\hat{\mathbf{w}}$.}
\label{fig:linear_fitting}
\end{figure}

The reasonable adaptation can again be seen with the correlations between the reconstructed populations in $\hat{\mathbf{w}}$ and the statevector values in $\mathbf{w}$, shown in Figure~\ref{fig:linear_fitting}. At $L=3$, the populations exhibit a strong linear correlation, with $R^2=0.98$, although the regression slope of $0.82$ indicates a systematic decrease in the population range. For $L=5$ and $7$, the correlations remain still substantial, with $R^2=0.90$ in both cases, while the slopes have decreased to $0.79$ and $0.67$, respectively. Only at $L=10$, $R^2$ falls to $0.19$ with a slope of $0.21$, indicating a pronounced loss in reconstruction quality. 

Taken together, these results show that noisy ancilla distributions retain useful information about the spin composition encoded by the target circuits with sufficient distinguishability and that our calibration scheme can detect it. The results also identify a practical limitation of increasing ansatz depth: additional layers can improve variational expressibility as seen in the preceding energy calculations shown in Figure~\ref{fig:graphs_asym}, while reducing the reliability of the measured spin composition. In fact, we admit that a subsequent VQD energy estimations based on $\mathbf{d}^\text{cal}$ (Eq. \ref{eq:d^cal}) failed to produce reliable potential energy surfaces. Thus, the present hardware benchmarks establish the feasibility and limitations of the reconstruction component paired with the presently available hardware. Even still, incorporating calibrated spin information into a complete shot-based sfVQD optimization is a further step, as it displays a future prospect with the steadily increasing hardware fidelity. Especially, already producing a reliable spin-sector distribution with the present day's hardware is indeed encouraging.

\subsection{Practical implications and broader interpretation}

Another practical advantage of our scheme is that it is obtained at the cost of only modestly increased circuit elements. One of the key parameters related to the cost is the number of the ancilla qubits $n_\mathrm{anc}$, and it scales logarithmically with the system size dictated by the numbers of electrons ($n_\text{elec}$) and spin-orbitals ($n_\text{sorb}$) as
\begin{equation}
    n_\mathrm{anc} \sim \log_{2}(\min(n_{\mathrm{elec}}, n_{\mathrm{sorb}}-n_{\mathrm{elec}})+1).
\end{equation}
This is because the ancillas should cover the entire range of possible angular momentum values that the Jordan-Wigner domain can take. For example, if there are four electrons in four spatial orbitals, up to quintet state is possible with the $m_{x}$ value ranging in $\{-2, -1, 0, 1, 2\}$, and three qubits are enough to express the five different possibilities. Thus, a small number of ancilla qubits are enough.
In addition, the added controlled-rotation layers remain shallow. Indeed, unlike approaches based on explicit $\hat{S}^2$ evaluation, the present scheme avoids substantially inflating the measurement structure associated with the Hamiltonian estimation. The protocol therefore remains compatible with NISQ-style resource constraints while providing improved spin selectivity.

The practical cost of spin screening also depends on the added operations applied to the system register after the electronic state has been prepared. On noisy hardware, this state is vulnerable to errors introduced by subsequent gate operations and by the additional time required for sensing. The screening circuit should therefore extract the required spin information with limited additional exposure to noise. Indeed, this consideration actually motivated the occupation-counting implementation used in our hardware experiments. Instead of performing repeated ancilla-controlled spin rotations with $\hat S_z$-based representation as in Figure~\ref{fig:R_{x}_screening}, we first transform the ansatz circuit to the $\hat S_x$ basis and then perform quantum additions separately on the $(+)$ and the $(-)$ domains to store the results to counting ancilla registers. (See Figure~S10 for its circuit diagram and the qubit topology this circuit benefits from.) Phase estimation is thus retained, but is applied to the number operators in the rotated basis. This construction provides a way to simplify the interactions involving the prepared electronic state while also making particle-number information accessible. Its practical benefit depends on the compiled gate sequence, connectivity, and execution time, rather than on the logical circuit structure alone.

The gain in spin purity, however, was not entirely free. In some cases, VQD/SSP exhibited slightly smoother energetic convergence than our sfVQD/SSP, as can be observed in Figure~\ref{fig:graphs_asym}. This behavior arose likely because the screening step modified the effective optimization landscape by suppressing contributions by shortcutting states that are wrong in their spin characters by themselves but are useful in connecting two other states with the correct spin property. Because the optimization with sfVQD walks through a more limited space with correct spin characters, a bypass through a wrong but shortcutting state is not allowed. Thus, optimization search with sfVQD may converge less gradually with some locally trapped features. The present results therefore reveal a practical trade-off between spin sector enforcement and optimization speed.

From a broader theoretical perspective, the present protocol is closely related to symmetry-based quantum sensing \cite{degenQuantumSensing2017, jamesMeasurementQubits2001}, in that it adopts controlled evolution under a symmetry generator to extract symmetry-sector information from the response of a quantum state. In our case, however, this information is not used as an end result in itself, but as a screening signal embedded within a variational excited-state workflow. When viewed in this way, our method can be regarded as a scheme with a symmetry-oriented sensing layer for state selection rather than as a direct $\hat{S}^2$ estimation scheme. This perspective also suggests possible extensions of quantum computing algorithms to utilizing other conserved quantities, such as molecular point-group irreducible representations or vibrational symmetries in highly symmetric molecules\cite{jangQuantumMechanicsChemistry2023}.
Indeed, one may treat a conserved quantity $Q$ by constructing a lightweight discriminator for $e^{i\theta Q}$ and incorporating its outcomes into a measurement--penalty or screening framework \cite{kitaevQuantumMeasurementsAbelian1995a, abramsQuantumAlgorithmProviding1999}. In the present work, total spin provides a stringent test case in that enforcing a fixed $m_{z}$ alone does not prevent spin contamination, while directly measuring $\langle\hat{S}^2\rangle$ typically incurs a substantial cost \cite{lyuSymmetryEnhancedVariational2023, sekiSymmetryadaptedVariationalQuantum2020}. Our results show that an $\hat{S}_{x}$-based discriminator can deliver robust spin-manifold separation with modest ancilla resources. Extending the same framework to larger active spaces and other symmetry constraints, such as point-group irreducible representations \cite{bunkerMolecularSymmetrySpectroscopy1998, cottonChemicalApplicationsGroup1990, setiaReducingQubitRequirements2020, jangQuantumMechanicsChemistry2023}, will be another natural direction.

\section{Conclusions}

We have introduced a spin-sensing strategy that combines a $\hat{S}_{z}$-conserving symmetry-preserving ansatz (SSP) with a shallow ancilla-assisted discriminator based on controlled rotations with $e^{i\theta\hat{S}_{x}}$, and demonstrated its application to spin-filtering variational quantum deflation (sfVQD). The SSP ansatz restricts the variational search to a manifold of fixed $m_{z}$, reducing the accessible Hilbert space size and the number of states that need to be orthogonalized in the VQD procedures. The $\hat{S}_{x}$-based screening module then acts as a probabilistic spin-space filter, allowing samples with ancilla readouts incompatible with the target spin sector to be rejected or penalized before Hamiltonian estimation. This suppresses spin contamination without explicitly evaluating $\langle\hat{S}^{2}\rangle$, with modest ancilla requirements. We have applied the scheme to calculating excited-state potential energy surfaces along symmetric and antisymmetric stretches of \ce{BeH2} and stretches of LiH, and demonstrated that sfVQD paired with the SSP ansatz markedly improves the separation of singlet and triplet state manifolds over conventional VQD with either SP or SSP ans\"atze. We further evaluated our spin-sensing component on noisy quantum hardware and found that our calibration scheme enables reconstruction of spin-sector populations to an extent that is modestly reliable even at the presently available noise level. The reconstruction became less reliable at larger ansatz depths, revealing a practical trade-off between variational expressibility and the accuracy of spin discrimination.

The hardware-based reconstruction was achieved using only 1024 shots per reference or target circuit, demonstrating that useful spin-sector information can be extracted with a modest sampling budget. Accurate energy estimation in VQE generally requires substantial sampling across Hamiltonian terms and optimization iterations \cite{tillyVariationalQuantumEigensolver2022}. Within such a workflow, allocating more shots to both calibration and target measurements may reduce statistical uncertainty and improve the precision of the reconstructed populations. The achievable improvement will depend on the distinguishability of the reference responses during calibration and the stability of the hardware, as additional sampling cannot remove systematic calibration errors.

More broadly, the present results support incorporating the measured effects of hardware noise into the design of quantum algorithms. The combination of symmetry-preserving state preparation, transverse-spin sensing, and calibrated reconstruction provides a way to use reproducible features of the noisy device response when extracting physically meaningful information. Within a stable circuit and hardware configuration, calibration references may also be reused across multiple target states, spreading their measurement cost over the variational search. Furthermore, the calibrated reference responses may support a statistical decision rule for spin screening. Although false rejections remain possible on noisy hardware, the rule could reject a trial state before all $N_{\mathrm{shot}}$ shots have been collected if the accumulated counts provide sufficient evidence that its off-target spin population exceeds a prescribed tolerance. This suggests that noise resilience can be pursued through the coordinated design of the circuit, measurement protocol, and classical inference procedure.

The accuracy of calibrated spin reconstruction is also expected to benefit from continued improvements in quantum hardware. It was reported that current Nighthawk processors exhibit $\sim10^{-3}$two-qubit gate errors and $\sim10^{-1}$ readout errors\cite{ibmCalibration}. More recently, Nighthawk r2 processor with improvements in qubit initialization and circuit throughput was announced\cite{haasNighthawkR2_2026}. Although quantitative determination on its improvements remains to be seen, reduced gate and readout errors are expected, which will surely enhance the spin-sector decompositions at greater circuit depths and enable more accurate population reconstruction with more reliable statistical screening, possibly also with reasonable estimations on the state energies.

\section*{Data availability}
The data that support the findings of this study are available from the corresponding author upon reasonable request.

\section*{Code availability}
The code used to generate the results reported in this study is available from the corresponding author upon reasonable request.

\section*{Acknowledgements}
This work was financially supported by the National Research Foundation of Korea (NRF) grants funded by the Korea government (MSIT), with Nos. RS-2020-NR049542, RS-2023-NR119931, and RS-2024-00432113. Administrational support by Creation of the Quantum Information Science R\&D Ecosystem (Based on Human Resources) Program through NRF (No. RS-2023-00256050) is also gratefully acknowledged in combination with RS-2025-08542968. Fruitful discussions with Jeyun Ju are also gratefully acknowledged. 

\section*{Author contributions}
Y.K.A. conceived of the project, developed the methodology, and performed all calculations with research supervision by Y.M.R. After Y.K.A. prepared its initial draft, both authors contributed to editing the manuscript.

\section*{Competing interests}
The authors declare no competing interests.

\bibliographystyle{naturemag}
\bibliography{SSP-ssVQD}
\end{document}